\documentclass[5p]{elsarticle}

\usepackage{lineno,hyperref}
\modulolinenumbers[6]

\usepackage[table]{xcolor}
\usepackage{amsthm}
\usepackage{amsmath}
\usepackage{graphicx}
\usepackage{tabularx}
\usepackage{subcaption}

\usepackage{multirow}
\usepackage{rotating}
\usepackage{booktabs}
\usepackage[final]{listings}
\usepackage{caption}
\usepackage{framed}
\usepackage[nounderscore]{syntax}
\newenvironment{BNF}
  {\captionsetup{type=lstlisting}}
  {}

\usepackage{geometry}
\usepackage{pdflscape}

\usepackage[many]{tcolorbox}
\newtcolorbox{framefloat}[1]{arc=0pt,outer arc=0pt,boxrule=0.5pt,
  colframe=black,colback=white,left=2mm,right=2mm}

\theoremstyle{definition}
\newtheorem{definition}{Definition}[section]

\journal{Journal of Systems and Software}

\begin{document}
\begin{frontmatter}

\title{Tigris: a DSL and Framework for Monitoring Software Systems at Runtime}

\author{Jhonny Mertz\corref{cor1}}
\ead{jmamertz@inf.ufrgs.br}
\cortext[cor1]{Corresponding author.}

\author{Ingrid Nunes}
\ead{ingridnunes@inf.ufrgs.br}

\address{Universidade Federal do Rio Grande do Sul (UFRGS), Instituto de Inform\'atica, Porto Alegre, Brazil}

\begin{abstract}
The understanding of the behavioral aspects of a software system is an essential enabler for many software engineering activities, such as adaptation. This involves collecting runtime data from the system so that it is possible to analyze the collected data to guide actions upon the system. Consequently, software monitoring imposes practical challenges because it is often done by intercepting the system execution and recording gathered information. Such monitoring may degrade the performance and disrupt the system execution to unacceptable levels. In this paper, we introduce a two-phase monitoring approach to support the monitoring step in adaptive systems. The first phase collects lightweight coarse-grained information and identifies relevant parts of the software that should be monitored in detail based on a provided domain-specific language. This language is informed by a systematic literature review. The second phase collects relevant and fine-grained information needed for deciding whether and how to adapt the managed system. Our approach is implemented as a framework, called Tigris, that can be seamlessly integrated into existing software systems to support monitoring-based activities. To validate our proposal, we instantiated Tigris to support an application-level caching approach, which adapts caching decisions of a software system at runtime to improve its performance.
\end{abstract}

\begin{keyword}
Monitoring \sep logging \sep execution trace \sep sampling \sep caching \sep performance
\end{keyword}

\end{frontmatter}
\linenumbers


\section{Introduction}

As modern software systems become increasingly large and complex, effective analysis methods to understand the dynamic behavior of a software system and act upon it in a timely fashion are becoming essential to ensure software quality. These analysis methods support fundamental software engineering tasks such as the addition of features, debugging, validation of quality requirements, performance engineering, or optimization~\cite{Feng2018}. Fully \emph{dynamic} and \emph{online} monitoring is the core of self-adaptive systems (SAS), which are those capable of adapting their behavior to keep satisfying their requirements in dynamic environments~\cite{Qin2016,Tanabe2017,Mertz2018a}. This typically involves collecting and storing runtime data from software components and the environment in such a way that the collected data can be analyzed and used to guide  adaptations in the system.

The system behavior is often recorded as \emph{execution traces}~\cite{Pirzadeh2011,Yuan2014,Reger2016}. This requires system instrumentation with additional code instructions that collect and store information from specific components of a system or its execution environment. As a consequence, there are practical challenges that must be overcome. First, the \emph{overhead} caused by the collection of execution traces in the observed system can become unacceptable, mainly for time-sensitive systems. Second, the result of system monitoring is a large set of traces, which frequently requires huge amounts of \emph{storage space}. These problems have been tackled on a case-by-case basis, e.g., by limiting the tracing only to high-level events~\cite{Horky2016,Kang2018} or a predefined set of executions~\cite{Pirzadeh2011,Miranskyy2016}. However, the former may be not enough depending on the purpose of monitoring, while the latter may not be known at design time or may require frequent revisions of the parts to be monitored. Consequently, fully monitoring the software system becomes the only viable alternative.

To provide a solution that can be reused across different systems and application domains, we propose a two-phase monitoring approach for filtering and sampling execution traces at runtime. In its first phase, a lightweight and coarse-grained monitoring is performed to identify relevant parts of the software execution. As relevance is essentially a domain-specific concept, this process is informed by the \emph{goal of monitoring} in the form of high-level \emph{relevance criteria} expressed in a proposed domain-specific language, called \emph{TigrisDSL}. In practice, the criteria are translated into software metrics, which are collected and analyzed at runtime to guide an in-depth and fine-grained monitoring in the second phase of the approach. Both relevance criteria and software metrics are derived from a systematic literature review.

Our approach is implemented as a framework, named Tigris, which seamlessly integrates the proposed solution to existing software systems to support monitoring-based activities. Consequently, our approach and framework can be used as a monitoring component to effectively monitor a software system and provide information for different purposes, e.g.\ to identify security vulnerabilities~\cite{Yuan2014}, model inconsistencies~\cite{Bartocci2018} or performance bugs~\cite{DellaToffola2015,Mertz2018a}. To validate our proposal, we instantiate Tigris as the monitoring component of an approach that improves the performance of applications using caching. Our evaluation shows that our proposal can maintain the effectiveness of the caching approach while reducing the monitoring overhead.

This paper is an extension of previous work~\cite{Mertz2019}. Compared to the original paper, it contains the following new contributions.

\begin{itemize}

    \item The foundational study from which we derived our relevance criteria was extended to include studies published in 2019.
        
    \item Our foundational study includes now the analysis of three additional aspects (scalability, generality and adaptivity) associated with monitoring-based ap\-proach\-es.
        
    \item Our preliminary version of the domain-specific language is now formalized as the \emph{TigrisDSL}, which allows the creation of monitoring filters by means of high-level relevance criteria.
    
    \item The two-phase monitoring approach, as well as its implementation as a framework, called Tigris, are explained in further detail and illustrated with a running example.
        
    \item The evaluation of the \emph{Tigris} framework has been extended with new research questions, exploring the adaptivity of the approach to variations in the workload mix, and different trade-offs between monitoring overhead and the reliability of the traces.
        
\end{itemize}

The remainder of the paper is organized as follows. Next, we discuss challenges and existing approaches in the context of software monitoring. In Section~\ref{sec:criteria}, we present a foundational study and its results, which includes the derivation of a domain-specific language, named TigrisDSL. Our proposed two-phase monitoring approach is introduced in Section~\ref{sec:proposal-overview}, and we give details of its implementation as a framework in Section~\ref{sec:framework-section}, An empirical evaluation of our approach is presented in Section~\ref{sec:evaluation}. Finally, we conclude in Section~\ref{sec:conclusion}.


\section{Challenges on Monitoring and Related Work}
\label{sec:related_work}

Runtime monitoring has been used for many purposes, from runtime verification to software self-adaptation~\cite{Wilhelm2008,Cornelissen2009,Derbeko2016,Rabiser2017}. We next further discuss the two major problems that must be addressed by adding monitoring at runtime and approaches proposed to address them.

\subsection{Reduction of the Monitoring Overhead}

A typical solution to reduce the monitoring overhead is to consider a sample of execution traces. The key advantage is the bounded overhead, which linearly decreases with the reduction of the sampling period and size. The most popular and straightforward sampling strategy is random or systematic sampling~\cite{Chan2003,Dugerdil2007}. However, this also results in a lower number of collected traces, which may be an inadequate representation of the population of traces. To address this, there are approaches that focus on sampling particular executed code locations and those that adjust the sampling rate.

\paragraph{Dynamic Monitoring Region}

There are approaches~\cite{Apiwattanapong2003,Santelices2006,Fei2006,Sridharan2007,Narayanappa2010} that monitor particular program regions or paths as the sampling strategy. However, this leads to a reduced coverage and possibly an unrepresentative sample~\cite{Pirzadeh2013}. Thomas et al.~\cite{Thomas2011} investigated several schemes for using markers to optimize linear sampling by reducing the sampling rate without increasing the original overhead, while Fischmeister and Ba~\cite{Fischmeister2010} built a system model and consider three theorems to determine the sampling period in different scenarios. Fei and Midkiff~\cite{Fei2006}, in turn, proposed a framework that avoids already monitored executions. This is based on the observation that the repeated execution of a (region of a) program with the same context tends to produce the same outcome. The use of these sampling or filtering schemes makes them suitable for monitoring-based activities that can tolerate some degree of data loss~\cite{Dwyer2007}.

\paragraph{Dynamic Monitoring Rate}

The monitoring overhead can be reduced by adjusting the sampling rate to the current system load. This leads to an adaptive sampling, which was proposed by Hauswirth and Chilimbi~\cite{Hauswirth2004}. The proposed sampling rate is inversely proportional to the frequency of the execution, ensuring the spatial coverage but not the temporal coverage of all program executions. This may, however, not be well suited to capture symptoms that are not uniformly distributed throughout the execution. Two approaches, in contrast, adjust not only the sampling rate but also its scope. Daoud et al.~\cite{Daoud2017} proposed a dynamic tracing approach based on conditions and thresholds, while Fei and Midkiff~\cite{Fei2006} proposed a runtime monitoring approach that balances the system load and the monitoring coverage. The latter serves for the specific purpose of identifying program regions where bugs are likely to occur.

\subsection{Management of the Size of Execution Traces}

The second problem associated with monitoring execution data is the amount of data generated to be stored and mined. In the face of large sets of traces, there is a need for huge storage space and analysis time, which may hinder opportunities for the timely detection of time-sensitive problems. Moreover, only a small subset of traces may be significant and enough for the analysis~\cite{Hamou-Lhadj2004,Pirzadeh2013,Miranskyy2016}.

In this sense, sampling-based filtering approaches have been proposed to cope with the size of execution traces by selecting a subset of traces based on event type, time, description, or even the priority or importance~\cite{Pirzadeh2013,Kouame2015}. Consequently, only the traces that match a particular pattern are selected~\cite{Zaidman2004,Reiss2005,Hamou-Lhadj2006,Cornelissen2008,Miranskyy2008,Pirzadeh2013}, such as being of a certain program region~\cite{Fei2006, Narayanappa2010}. Although these approaches manage the size problem, it is not always trivial to specify monitoring patterns and triggers to obtain traces of interest. Furthermore, these specifications are usually fixed, which limits the flexibility and adaptability of the monitoring and couples the tracing strategy with the analysis, resulting in a possibly hidden relevant behavior that remains undetected~\cite{Horky2016}.

\subsection{Discussion}

Runtime monitoring often leads to an overhead in the observed program, being mostly used within program analysis tools and having limited practical use to change or adapt program behavior at runtime~\cite{Kang2018}. In the latter case, the monitoring is often limited to logging high-level events. Detailed measurements, e.g.\ method-level tracing, tend to be avoided because their overhead can disrupt execution. This limits the information available for analysis when finding and solving issues at the code level~\cite{Horky2016}.

Filtering and sampling has been demonstrated as a practical solution to reduce both the monitoring and trace size issues. However, the majority of the existing solutions are conceived to be used in a post-mortem manner, which does not consider the monitoring overhead. In addition, low-overhead runtime monitoring strategies serve for specific purposes and, consequently, are not generalizable.

Although there are generic frameworks~\cite{Wininger2017,Hassine2018} to analyze, comprehend and visualize execution traces, they focus on debugging, visualization and comprehension of the software system behavior, demanding manual analysis to investigate and solve issues. In addition, these approaches require the manual tuning of several parameters, e.g.~\cite{Kouame2015} and~\cite{Eichelberger2014}, which consist of approaches that demand the definition of custom filtering behavior patterns in execution traces through several parameters and configuration files. Finding the adequate parameters can be a difficult task, and effective only for specific sets of traces (even within a single system).

Thus, there is a need for monitoring techniques that (1) make an adequate trade-off between the monitoring overhead, by not compromising the system operation and still collecting relevant execution traces; and (2) can be used for different purposes, such as detecting failure points or performance issues. We take a step towards this by proposing a generic solution for collecting relevant execution traces according to the goal of monitoring and analysis. It uses specified high-level relevance criteria, introduced next.

\section{Relevance Criteria}
\label{sec:criteria}

A complete set of execution traces captures the execution of every operation of a software system. However, not every trace is equally relevant given a particular goal when monitoring the system, because only a subset of the traces contains the information needed to diagnose a target system symptom. These traces may be concentrated in parts of the code with specific characteristics. For example, if the goal of the monitoring is runtime verification~\cite{Hamou-Lhadj2004,Reger2016}, methods that handle many exceptions or include type castings may be the primary source of relevant traces, as such methods are error-prone. Alternatively, if the purpose of monitoring is performance optimization, methods that are more frequently invoked might be those to be tracked~\cite{Chen2018}. The relevance of an execution trace for analysis depends on the \emph{purpose} of the analysis, i.e.\ the monitoring goal. A sample of collected execution traces is said \emph{relevant} if the portions of the source code being monitored satisfy a set of relevance criteria. A relevance criterion is defined as follows.

\begin{definition}{\emph{Relevance Criterion.}}
A relevance criterion is the specification of a property of system events (e.g.\ the execution of a method or a function) that characterizes the types of events that are more likely to be useful than others, according to a particular monitoring goal.
\end{definition}

To identify the relevance criteria associated with each monitoring goal, we surveyed monitoring-based approaches from the literature, identifying their goals as well as adopted criteria and metrics to analyze execution traces to understand the system behavior. Next, we first describe the method used to select and analyze research work in this context and then present obtained results. Founded on them, we introduce the \emph{TigrisDSL}, which is a domain-specific language (DSL) to specify relevance criteria.

\subsection{Method}\label{subsubsec:method} 

We surveyed full papers published in the Computer Science conferences presented in Table~\ref{tab:conferences}. These conferences have on their scope of interest research related to the analysis of execution traces for profiling, adaptation or code understanding. We selected the two most highly ranked conferences in software engineering (ICSE and ESEC/FSE), and five other conferences where system monitoring is a core concern due to its importance to the software adaptation area (ICAC, SASO, and SEAMS) and code analysis and understanding (OOPSLA and ICPC). We focused only on conferences to pursue face-paced publications considering that they have faster review and publication processes than journals. Moreover, papers published in journals are often extensions of conference papers. In addition, we also covered journal-first publications of relevant journals in the area, as these papers can be presented at selected conferences. We restricted our search to the past six years (2014--2019). To filter relevant papers, we searched the databases where the proceedings of these conferences were published using the following query string.

\begin{table*}[!t]
\centering
\caption{Conferences from where papers from the past six years (2014--2019) were obtained and analyzed.\label{tab:conferences}}
    \begin{tabularx}{\linewidth}{l X}
        \toprule
        \textbf{Acronym} & \textbf{Conference} \\ \midrule
        ESEC/FSE & Joint European Software Engineering Conference and Symposium on the Foundations of Software Engineering \\
        ICSE & International Conference on Software Engineering \\
        OOPSLA & Conference on Object-Oriented Programming Systems, Languages, and Applications \\
        ICAC & International Conference on Autonomic Computing \\
        SASO & International Conference on Self-Adaptive and Self-Organizing Systems \\
        SEAMS & International Symposium on Software Engineering for Adaptive and Self-Managing Systems \\
        ICPC & International Conference on Program Comprehension \\ \bottomrule
        \end{tabularx}
\end{table*}

\begin{framed}
\noindent (\emph{runtime} OR \emph{dynamic}) AND (\emph{monitor} OR \emph{instrument} OR \emph{record} OR \emph{track} OR \emph{profile}) AND (\emph{trace} OR \emph{execution})
\end{framed}

As result, we obtained 147 papers, from which 64---based on the title and abstract---fit the criterion of including a monitoring-based approach. They were analyzed according to four dimensions:

\begin{enumerate} 
    \item identification of the monitoring \emph{goal}, the \emph{criteria} adopted to analyze execution traces and the used \emph{metrics}; 
    \item assessment of the \emph{generality} of the monitoring approach in terms of its use for different application domains and purposes;
    \item examination of \emph{scalability} aspects, such as the overhead and size of traces; and 
    \item inspection of the \emph{adaptability} of the monitoring approach, that is, the runtime changes in the monitoring strategy to address a particular issue, such as the reliability of collected traces or overhead reduction. 
\end{enumerate}

To guide the analysis of the selected papers, we used pre-specified questions, presented in Table~\ref{tab:analysis_questions}, which served as a checklist while analyzing the papers. The analysis was performed manually by reading the selected papers and answering the questions. Table~\ref{tab:analysis_questions2} provides examples of such analytical processes and illustrates how we reached the results of our study\footnote{The detailed analysis of each paper is available at \url{http://www.inf.ufrgs.br/prosoft/resources/2020/jss-effective-tracing}}.

\begin{table}[!t]
    \caption{Analysis Approach. Questions used in the analysis of monitoring-based approaches.\label{tab:analysis_questions}}
        \begin{tabularx}{\linewidth}{l X}
        \toprule
        \textbf{\#Q} & \textbf{Question} \\ \midrule
        1 & What is monitored? \\
        2 & What are the metrics used? \\
        3 & What are the criteria related to the metrics? \\
        4 & What is the goal of monitoring? \\
        5 & How are traces collected? \\
        6 & What is the granularity of the collected traces? \\
        7 & What is the amount of the collected traces? \\
        8 & How were the traces generated for evaluation? \\
        9 & Is the monitoring automatically integrated? \\
        10 & Is the monitoring adaptive? What is the goal of adaptation? \\
        11 & How is adaptation achieved? What is the adaptation trigger?	\\
        12 & Is there monitoring overhead? How is the overhead dealt with? \\
        13 & Is representativeness considered when collecting traces? \\
        14 & How is representativeness ensured? \\
        15 & Is relevance considered when collecting traces? \\
        16 & How is relevance ensured? \\
        17 & How specific is the monitoring approach in terms of technology or domain/purpose? \\ \bottomrule
        \end{tabularx}
\end{table}

\begin{table*}[!t]
    \caption{Analysis Approach. Example of how information from the selected papers were collected by answering the questions specified in Table~\ref{tab:analysis_questions}.\label{tab:analysis_questions2}}
        \begin{tabularx}{\linewidth}{X l l}
        \toprule
        \textbf{Sample Quote from Selected Papers} & \textbf{Gathered Information} & \textbf{\#Q} \\ \midrule
        \cite{Barna2017}: The proposed monitoring module gathers CPU utilization for all \textbf{the back-end tiers of the applications (Web, Spark and Cassandra) on the container level}, CPU utilization of \textbf{the host VMs}, as well as response time and throughput from \textbf{the entire system}. & Target of monitoring & 1 \\ \midrule
        
        \cite{DellaToffola2015}: [...] uses state of the art CPU time profiling to identify the set of initial memoization candidates. We record, \textbf{for each executed method \textit{m}, the time \textit{$t_{m}$} spent in \textit{m}} (including time spent in callees) and \textbf{the number \textit{$c_{m}$} of calls}. Furthermore, we also measure \textbf{the total execution time \textit{$t_{prgm}$}} of the program. & Metrics & 2 \\ \midrule
        
        \cite{Kang2016}: \textbf{Based on the profiling data, DiagDroid detects performance issues and analyzes their causes}. A report can finally be generated with an \textbf{aim to direct the debugging process}. & Goal of monitoring & 4 \\ \midrule
        
        \cite{Su2016a}: There are several factors that can contribute to the \textbf{runtime overhead of HitoshiIO} [...] The most dominant factor for execution time in our experiments was the clone identification time: application analysis was relative quick (order of seconds), and the input-output recorder \textbf{added only a roughly 10x overhead} & Scalability & 12 \\
        \bottomrule
        \end{tabularx}
\end{table*}

The combined analysis of the four dimensions based on information gathered from the selected papers provided us with the foundation needed to elaborate a solution that can effectively support the development of monitoring-based approaches. The results are presented as follows.

\subsection{Analysis and Results}\label{subsubsec:results}

\subsubsection{Goals, criteria, and metrics}

Based on the analysis of our selected papers, we identified the criteria presented in Table~\ref{tab:relevance_criteria}. This table also shows the number of papers in which each criterion appears. As can be seen, the \emph{frequency} of an execution event is the most adopted criterion to classify relevant traces, followed by \emph{maintainability}.

\begin{table}[!t]
    \caption{Relevance Criteria. List of relevance criteria identified in our searched literature, with their description. The number of occurrences (\#) in the analyzed papers is also detailed.\label{tab:relevance_criteria}}

        \begin{tabularx}{\linewidth}{l p{4.5cm} r}
        \toprule
        \textbf{Criterion} & \textbf{Description} & \textbf{\#} \\ \midrule
        Frequency & Amount of times that an event occurs during a period of time &  30 \\
        Maintainability & Complexity of the operations associated with an event & 20 \\
        Expensiveness & Consumption of computational resources associated with an event & 16 \\
        Changeability & Analysis of divergences among the result of multiple occurrences of a same event & 14 \\
        Error-proneness & Likelihood of an event to cause errors when they occur & 13 \\
        Usage pattern & Characteristics of an event that tracks or deals with user requests & 12 \\
        State variation & Amount of changes in the system state caused by an event & 9 \\
        Concurrency & Amount of execution threads that are executed in parallel and share resources & 8 \\
        Latency & The delay of the occurrence of an event & 8 \\ \bottomrule
        \end{tabularx}
\end{table}

Although the order in this table may give evidence of the importance of the criteria, note that each criterion has a different purpose. Consequently, the number of occurrences is also associated with the most common goals of monitoring. We thus, in addition to the identification of criteria, investigated the goal of monitoring in the selected papers. The identified goals are clustered into five groups, as presented in Table~\ref{tab:relevance_criteria_metrics}. For each group, we show examples of specific goals that appeared in the papers. We also highlight the most used criteria for each goal. To assess whether execution events match the relevance criteria of a specific goal, various metrics are used to make an objective evaluation. Cells of Table~\ref{tab:relevance_criteria_metrics} indicate the most relevant metrics associated with each pair of goal and criterion that appeared in our surveyed literature.

\begin{table*}[!t]
        \centering
        \caption{Goals, Relevance Criteria and Metrics. Association between groups of goals and relevance criteria, along with examples of goals and metrics used by the surveyed approaches. Cells in gray highlight the three most relevant criteria of each goal. The given examples are non-exhaustive. \label{tab:relevance_criteria_metrics}}       
\begin{sideways}
    \begin{tabular}{llp{3.3cm}p{3.3cm}p{3.3cm}p{3.3cm}p{3.3cm}}
        \toprule
         \multicolumn{2}{l}{\textbf{Goal Group}} & Efficiency & Maintainability & Reliability & Security & Testability \\ \hline 
         \multicolumn{2}{l}{\textbf{Goal Examples}} & performance, energy saving, caching, improving resource consumption & bug finding, understanding, reuse, documentation, troubleshooting & health checking, fault tolerance, disaster recovery, adaptation, configuration fix & anomaly detection, data protection, malicious attack detection & testing (generation, validation, selection), reporting, verification \\ \midrule
         & Frequency & \cellcolor{gray!25} Number of occurrences in a period & \cellcolor{gray!25} Number of references and dependencies & \cellcolor{gray!25} Inter-arrival times & Changes in occurrence history & \cellcolor{gray!25} Number of occurrences in test case \\ \cmidrule(l){2-7} 
         & Maintainability & Number of operations involved & \cellcolor{gray!25} Static source code metrics & --- & Contextual information of objects/classes & \cellcolor{gray!25} Fail test coverage \\ \cmidrule(l){2-7}
         \multicolumn{1}{l}{\multirow{11}{*}{\rotatebox[origin=c]{90}{\textbf{Criterion}}}} & Expensiveness & \cellcolor{gray!25} Execution time & Source code locations of expensive methods & \cellcolor{gray!25} CPU and heap utilization, processing times & \cellcolor{gray!25} Transaction duration & Depth of call stack \\ \cmidrule(l){2-7} 
         & Changeability & Number of repeated computations & Similarity between call graphs & Number of operations with cached results & \cellcolor{gray!25} Changes in contextual information & --- \\ \cmidrule(l){2-7} 
         & Error-proneness & Number of failures of a component & \cellcolor{gray!25} Number of handled exceptions & \cellcolor{gray!25} Number of failures perceived by users & Increase of failures in a specific component & \cellcolor{gray!25} Number of failure assertions or exception-throwing statements \\ \cmidrule(l){2-7} 
         & Usage pattern & Changes in user navigational activity & --- & Number of active users and idle/active intervals & \cellcolor{gray!25} Variations in the request payload for same operations & --- \\ \cmidrule(l){2-7} 
         & State variation & I/O consumption per operation & Changes in the system state & Number of write operations performed & --- & --- \\ \cmidrule(l){2-7} 
         & Concurrency & Number of active users and threads & Number of references and dependencies & Number of race conditions & --- & Number of locks per test case \\ \cmidrule(l){2-7} 
         & Latency & \cellcolor{gray!25} Processing and bandwidth consumption & --- & Throughput & --- & --- \\ \bottomrule 
    \end{tabular}
\end{sideways}
\end{table*}

\subsubsection{Generality} 

All the surveyed approaches have been proposed for a specific goal, such as identifying performance bottlenecks or bugs. Despite their sheer number and heterogeneity, 49 papers (76.5\%) are limited to specific purposes, individual systems, particular architectural styles, or technologies, which couples the monitoring phase of these approaches to the analysis they perform (i.e.\ the goal of monitoring). These solutions are limited in terms of reuse because they are developed in an ad-hoc manner and require re-engineering work in order to adapt applications to obtaining tracing features.

From the surveyed approaches, 15 papers (23.5\%) are flexible in terms of configuration. It means that they can be customized in terms of constraints, rules, and properties of the approach to better fit the user-specific needs. This flexibility is achieved by offering lower-level interfaces, functions, and probes that can be manually implemented and customized by users, such as in~\cite{Jung2014,Lee2016,Devries2018}. However, only 5 papers (7.8\%) increase the \emph{generality} of monitoring by providing higher-level support so that users can achieve a domain-specific specification of the goal of monitoring, which is automatically implemented by the monitoring proposal. This is offered in the form of annotations, parameterization or domain-specific languages, e.g.\ in~\cite{Ghezzi2014,Angelopoulos2016,Christakis2017}.

In addition to analyzing whether the monitoring can be customized, we also inspected if it relies on the particularities of specific technologies, programming languages, or execution environments. From the 64 papers, only 17 (26.5\%) are technology-independent. The remaining 47 surveyed approaches (73.5\%) rely on traces and events with properties and format tight to specific programming languages such as JavaScript~\cite{Gong2015} or Java~\cite{Huang2015}. The same applies to approaches focused on software platforms such as Android~\cite{Kang2016} and Linux~\cite{Song2017}. In these scenarios, traces are usually collected by instrumenting or profiling the execution platforms at a lower-level, and thus the monitoring approach becomes dependent on the platform specificities.

This means that most of the existing solutions that rely on monitoring are application-specific, i.e.\ while they carefully collect selected data at runtime for a particular purpose, the principles used to monitor data are not generalizable. However, the rationale behind the criteria used to filter traces is shared by the approaches.

\subsubsection{Scalability} 

Monitoring and collecting information from a managed system impact on its performance. From all  approaches, 31 papers (48.5\%) explicitly mention that their solution implies an overhead to the observed system, such as in~\cite{Su2019,Bronink2016,Madsen2016,Zhang2014,Bielik2015,Kantert2014}. The overhead implied may vary according to the amount of information required to the analysis and the monitoring technique used to collect the execution traces. For example, Su et al.~\cite{Su2016a} report an overhead of 10$\times$ compared to running a Java-based application without the profiling technique used to detect functional clones, which relies on collecting detailed inputs and outputs of function calls. A different monitoring technique adopted by Hu et al.~\cite{He2018}, which relies on collecting occurrences of operational system calls based on a Linux kernel (low-level) extension, reports an overhead of less than 1\%. The remaining 33 surveyed approaches (51.5\%), e.g.~\cite{Camara2014,Tun2018,Bastani2015,Larsson2019,Chen2016,Denaro2015}, do not even mention the impact of monitoring and gathering data from the managed system. This could raise questions of their practical feasibility because even in cases where a small amount of data is collected during system monitoring, like execution time and identification of the event~\cite{He2018,Yang2019}, there is an impact on the memory consumption and processing time of the system.

Sampling and filtering have been demonstrated as the most used solution to reduce the trace size and enable faster trace analysis, adopted by 26 approaches (40.5\%). However, the reduction of the amount of traces may lead to an unrepresentative sample and, consequently, inadequate to achieve the goal of monitoring. None of the surveyed approaches ensured the representativeness of the sample collected.

In terms of overhead evaluation, only 16 papers (25\%) assessed the extent of the overhead. The reported overhead varies from negligible (0.8\%--5\%)~\cite{Jung2014,Kim2016a,Kang2016,Hawkins2017,Lee2014,He2018,Liu2016} to high impact (such as +16\% or 41.7$\times$ the original execution)~\cite{Su2019,Aliabadi2017,Bronink2016,Jensen2015,Gong2015,Madsen2016,Zhang2014,Bielik2015,Su2016a}, which compromises the practical feasibility of the approach in real-time scenarios. These high impact approaches are typically not supposed to be used at runtime because they demand detailed monitoring of the system and are usually focused on testing scenarios, possibly with production workloads. The results can be manually analyzed and applied by developers at design time to benefit the system in future executions.

Towards addressing the observed monitoring overhead, 16 of the approaches (25\%) explored alternatives to reduce the impact of the monitoring activity. From these, six~\cite{Jensen2015,Madsen2016,Huang2015,Samak2016,Bielik2015,Xiao2014} apply specific tuning and optimization in the proposed algorithms to reduce the overhead, which is not possible to generalize to other approaches. For example, Madsen et al.~\cite{Madsen2016}, who focused on providing developers with detailed information about crashes at execution time, iteratively increase the instrumentation level on regions of code in which crashes are detected. In addition, Huang et al.~\cite{Huang2015}, which monitors thread-related operations in Java-based applications, do not collect global traces but instead focus on the events of each thread separately. The remaining ten approaches~\cite{Su2019,Kim2016a,Bronink2016,Gong2015,Song2017,Chen2014,Jung2014,Lee2014,DellaToffola2015,Zhou2016} make use of generic solutions such as sampling (simple systematic or random) and static analysis to filter out locations and avoid collecting the so-called useless traces.

The granularity level of the monitoring may play an important role in the overhead. In this regard, 40 of the surveyed approaches (62.5\%) rely on monitoring function or method calls, which implies a considerable overhead given the high number of traces and the detailed information usually collected from method calls such as the input parameters and return. In 36 approaches (56\%), the trace collection is performed based on low-level profiling and instrumentation, according to the programming language in which the approach is implemented such as based on Low Level Virtual Machine\footnote{\url{https://llvm.org/}} (LLVM)~\cite{Song2017,Kim2016b} for C++ applications, Jalangi framework\footnote{\url{https://jacksongl.github.io/files/demo/jalangiff/index.html}}~\cite{Gong2015,Jensen2015} for Javascript-based applications, or ASM-based\footnote{\url{https://asm.ow2.io/}} instrumentation~\cite{DellaToffola2015,Zhang2014,Chen2018,Su2016a} for Java applications. In addition, eight approaches (12.5\%) either demand manual implementation of the tracing collection code~\cite{Angelopoulos2016,Ghezzi2014,Yandrapally2015,Xu2016,Su2016b} or provide ways of generating the implementation automatically~\cite{Lee2014,Xiao2014,Devries2018}. However, code-level changes imply maintenance issues and increase the complexity of the software system base code.

For practical reasons, sometimes the monitoring is limited to high-level events~\cite{Camara2014,Clark2015,Barna2017,Yuan2014,Brocanelli2017,Dong2018,Zhou2019}. Examples of this type of event are requests to a web server~\cite{Camara2014} or occurrences of failures in software components~\cite{Zhou2019}. This is the case of 15 of the surveyed approaches (23.5\%) that rely on system or module-level monitoring. However, this may reduce the power of the analysis, given the lower amount of collected information.

\subsubsection{Adaptability}

The way that the monitoring is performed can change at runtime, either to focus on relevant traces or to reduce the overhead. Examples of changes are the metrics being collected, sampling configurations, or target locations oriented by a domain-specific analysis and occurrence of an event of interest. These are forms of adaptive monitoring.

Only seven of all surveyed approaches (11\%) employ adaptive mechanisms to improve the monitoring efficiency~
\cite{Bronink2016,Gong2015,Madsen2016,Samak2016,DellaToffola2015,Barna2015,Casanova2014}. In all cases, the adaptation is triggered based on a set of specific constraints or hard-coded rules and are thus not flexible to be modified. For example, Bronik et al.~\cite{Bronink2016} increase the reliability of collected data by dynamically placing probes in component connections according to fault detection. As its goal is fault localization, it changes the trace rate to have more information about which components are more susceptible to faults. It triggers the adaptation based on rules such as if a problem has been diagnosed in a component, probes are deployed to obtain a more accurate diagnosis.

Another example is the evaluation of the health of a managed resource~\cite{Gong2015}. Because the instances in a distributed system can come and go dynamically, when a new component appears or leaves the system, the monitoring component is capable of reflecting changes in the topology and keeping the reliability of the collected data. Adaptation is also used to dynamically reduce the level of the overhead. Barna et al.~\cite{Barna2015}, e.g., adapt the sampling technique to focus on specific code locations depending on the computational resources available for monitoring.

In addition, all the proposed adaptations are focused on controlling the monitoring overhead~\cite{Barna2015,Madsen2016,Samak2016}, or increasing the trace reliability for a specific purpose~\cite{Bronink2016,Gong2015,Casanova2014,DellaToffola2015}. All of them are limited to changing configurations based on pre-defined setups or ad hoc assumptions, without considering the trade-off involved in the process. An example of such an assumption is that the performance overhead is reduced by merely reducing the target locations~\cite{Barna2015}, which may not be valid if the majority of the system executions are concentrated on the filtered locations. Thus, generic adaptation goals, such as ensuring the representativeness of the collected traces or dynamically managing the monitoring overhead, are not addressed by any of the surveyed approaches.

\subsubsection{Discussion}

Although runtime monitoring approaches have been employed to different goals and purposes, there are still limitations in terms of generality, scalability, and adaptability that should be addressed in order to achieve an effective monitoring approach. Thus, there is a need for a monitoring technique that:

\begin{enumerate}

    \item can be applied generically and flexibly for different types of software systems and purposes, such as detecting and dealing with performance issues or energy bottlenecks, considering different levels of monitoring granularity;
    
    \item can deal with the trade-off between the impact caused by the monitoring and its effectiveness in terms of data representativeness, relevance and location coverage; and
    
    \item is able to respond to changing requirements and constraints in the monitoring component in order to maintain the monitoring effectiveness.
    
\end{enumerate}

In addition, the development of reusable monitoring approaches that abstract and encapsulate monitoring functionality would reduce both the effort to develop new systems implementing these strategies as well as the probability of bugs in newly implemented solutions. Moreover, it would promote software reuse across different goals and domains. Based on the findings of our survey of monitoring-based approaches, we derived \emph{TigrisDSL}, a do\-main-specific language (DSL) designed to provide users with a standardized and comprehensive way to specify monitoring goals in terms of metrics and relevance criteria. TigrisDSL is founded on the relevance criteria presented in Tables~\ref{tab:relevance_criteria} and~\ref{tab:relevance_criteria_metrics} and is the basis of our monitoring solution (introduced in Section~\ref{sec:proposal-overview}).

\subsection{TigrisDSL: a Generic Way to Specify Relevance Criteria}\label{subsec:dsl}

TigrisDSL allows users to write monitoring filters by means of high-level relevance criteria. These relevance filters can be used to guide monitoring components to collect a set of relevant traces that are analyzed to achieve the goal of monitoring. It can be incorporated into any monitoring approach to specify monitoring requirements.

The Backus–Naur Form (BNF) grammar of TigrisDSL is presented in Listing~\ref{grammar:dsl}. TigrisDSL is based on the recursive definition of a \texttt{filter}, which can be composed of multiple definitions (\texttt{filterdef}). Filter definitions are the main structure to allow users to inform which group of data from the relevance criteria should be considered (\texttt{modifier}) and the relevance criteria from the set of pre-defined criteria (\texttt{criterion}). It is important to highlight that these pre-defined criteria are based on our systematic literature review. Nevertheless, future versions of our DSL can include extended criteria or modifiers if those derived from our systematic analysis are considered not enough for the specifications of filters.

The semantics of the \texttt{modifier} should be specified by an approach employing our DSL. For example, \texttt{frequent} can be events that are executed in a frequency above the average, while \texttt{most frequent} can be the top 5\% of the most frequent execution events (e.g.\ invoked methods or called functions).

\newpage

\begin{BNF}
\caption{TigrisDSL Syntax Grammar. Presentation of the BNF grammar of the TigrisDSL language.\label{grammar:dsl}}
\begin{framed}
\begin{grammar}
<filter> ::= <filterdef> | ( <filter> ) | <filter> <operator> <filter>

<filterdef>  ::= <modifier> <criterion> | <criterion>

<operator> ::= $\cup$ | $\cap$ | $\setminus$

<modifier> ::= more | less | most | least

<criterion> ::= frequent | maintainable | expensive | changeable | error-prone | usage-pattern | state-variation | concurrent | latent
\end{grammar}
\end{framed}
\end{BNF}

In addition, \texttt{filter}s can be combined to form a complex filter using \texttt{operator}s, which are based on basic \texttt{set operations}, namely union, intersection, and subtraction. With these operations, it is possible to specify how the data from different groups of the relevance criteria can be combined to identify and filter a set of relevant events. To illustrate, we give examples of expressions written in TigrisDSL as follows.

\begin{itemize}
    \item \texttt{least frequent}, which indicates that in terms of frequency, only the least frequent events of the system execution should be traced.
    \item \texttt{more frequent $\cup$ most expensive}, which indicates that the monitoring should be focused on methods that are more frequent or most expensive, considering all the system events.
    \item \raggedright\texttt{most changeable $\cap$ (most concurrent $\cup$ more error-prone)}, which indicates that only the most changeable events, which also have higher levels of concurrency or tend to cause errors, should be traced.
\end{itemize}

Depending on the semantics of \texttt{modifier}s, increasingly complex expressions can be specified as needed. Examples are presented as follows.

\begin{itemize}
    \item \raggedright\texttt{less changeable $\cap$ more frequent $\cap$ (more usage-pattern $\cup$ (more expensive $\cap$ less usage-pattern))}
    \item \raggedright\texttt{(least changeable $\cup$ most changeable) $\cap$ more frequent $\cap$ (most usage-pattern $\setminus$ less expensive)}
\end{itemize}

In order to demonstrate how relevance criteria can be used to abstract the desired behavior and filter execution traces for a specific purpose, we take a monitoring-based approach as an example~\cite{DellaToffola2015}. This approach was identified in our literature survey. Della Toffola et al.~\cite{DellaToffola2015} proposed a method to identify memoization opportunities based on profiling method executions of applications. During this analysis, all the method calls are filtered by processing three specifications, which are presented in the first column of Table~\ref{tab:running-example}. These specifications are informal and presented in a non-standardized way. The other two columns of this table show with which relevance criterion each specification is associated and a filter that represents it. The filters shown in the third column of Table~\ref{tab:running-example} are less ambiguous and more concise than natural language. In addition, they are a generic and explicit way to express what sort of monitoring events of interest.

\begin{table*}
    \caption{Relevance Filter Examples. Example of a monitoring-based approaches specified in TigrisDSL.\label{tab:running-example}}    
        \begin{tabularx}{\linewidth}{X l l}
        \toprule
        \textbf{Specification made by Della Toffola et al.~\cite{DellaToffola2015}} & \textbf{Relevance} & \textbf{TigrisDSL-based} \\ 
         & \textbf{Criterion} & \textbf{Filter} \\ \midrule
        The program spends a non-negligible amount of time to process a method. & Expensiveness & \texttt{more expensive} \\ \hline
        The program repeatedly provides equivalent inputs to a method, and this method repeatedly produces the same outputs for these inputs. & Changeability & \texttt{$\cap$ least changeable} \\ \hline
        The number of times that a result can be reused over the total number of cache lookups is at least 50\%. & Frequency & \texttt{$\cap$ more frequent} \\ \bottomrule
        \end{tabularx}
\end{table*}

It is important to highlight that the proposed language TigrisDSL is generic and abstract in the sense that it does not define the semantics of criteria such as frequency or expensiveness, as well as the meaning of more or less error-prone when comparing execution traces. Essentially, TigrisDSL captures the most representative concerns about the monitoring observed in the analyzed papers (criteria) and provides a syntactic construction to express a comparison among elements within a criterion in quality or degree (modifiers), along with a way to correlate and operate on top of different criteria to create relevance filters (operators). The filter expressions made using TigrisDSL can be used as input of any monitoring approach. To use TigrisDSL to create event filters, it is necessary to employ a mechanism that can interpret and translate these criteria, modifiers, and operators into quantitative and comparable metrics. We, in particular, propose a two-phase monitoring approach, which is guided by user-made specifications using TigrisDSL and provides semantics to the language criteria, modifiers and operators. Our proposal is a step towards achieving an effective monitoring approach and provides a way to define and customize monitoring components.


\section{A Two-Phase Approach for Collecting Execution Traces}
\label{sec:proposal-overview}

As discussed, monitoring all execution traces in detail comes at the cost of extensive and detailed instrumentation, which causes a high overhead in software applications~\cite{Mertz2017b}. Moreover, there are situations when it is infeasible to select software locations or executions that should be monitored a priori in detail (i.e.\ design-time), or such locations frequently change overtime. It thus becomes necessary to rely on an automated and adaptive process that can identify such executions of interest (e.g.\ the most frequent or more error-prone executions) with reduced performance overhead. Available monitoring approaches in this direction lack generality and fail in enabling software reuse across different domains with varying goals.

\begin{figure}
\centering
\includegraphics[width=\linewidth]{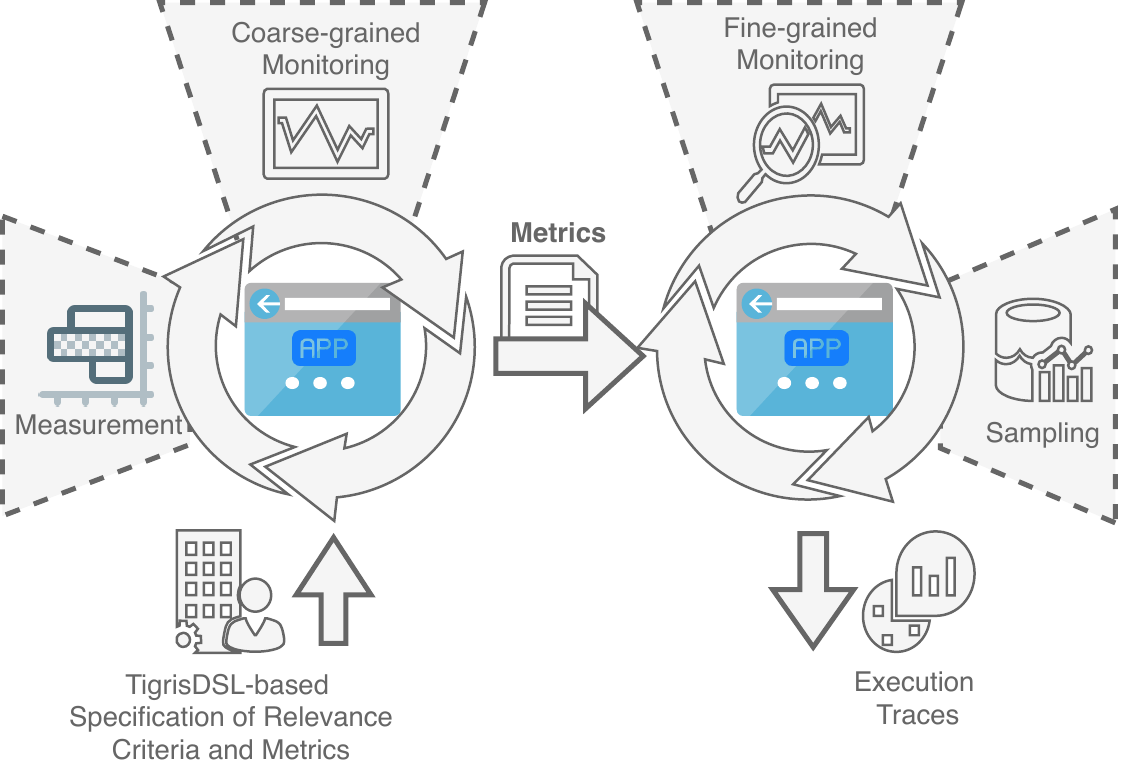}
\caption{Overview of the Two-phase Monitoring Approach. It shows the input provided in the first phase (Coarse-grained Monitoring) and the resulting output of the second phase (Fine-grained Monitoring).}
\label{fig:abstract_solution}
\end{figure}

To address this, we propose a two-phase monitoring approach for collecting execution traces, which is a generic and customizable solution for monitoring. As presented in Figure~\ref{fig:abstract_solution}, our approach is driven by user definitions supported by the proposed TigrisDSL, and thus provides semantics to all the terms of the TigrisDSL, such as criteria, modifiers, and operators. The first monitoring phase (described in Section~\ref{sec:coarse-grained}) is \emph{coarse grained}, focused on computing low-overhead metrics of event executions according to the specification of relevance criteria. The second phase (detailed in Section~\ref{sec:fine-grained}) takes as input the data collected in the coarse-grained phase. It processes the computed metrics from the previous phase and dynamically identifies the relevant events that are assumed to generate traces that are relevant for a given goal. These traces are collected in detail by a monitoring process that relies on a sampling strategy to control the overhead of monitoring. Before detailing the phases of our monitoring approach, we next present a running example that is used throughout this section to illustrate details on how the proposed approach works.

\subsection{Running Example: Application-level Caching}
\label{subsec:running-example}

Previous work~\cite{Mertz2018a} proposed an automated caching approach, which chooses and manages cacheable content according to the \emph{Cacheability Pattern}~\cite{Mertz2017a}. It targets the caching of \emph{method-level content}. The automatically selected cache configuration is based on observations made by monitoring web applications at runtime, that is, a monitored application workload. This approach was conceived and implemented in the form of a caching framework, named \emph{APLCache}, which seamlessly integrates the proposed solution to web applications.

One of the limitations of APLCache is that the overhead of the data tracking activity may significantly impact the application execution because it is necessary to monitor method inputs and outputs to make caching decisions. This was addressed by disabling the monitoring in situations when it is not possible to keep an acceptable overhead. Thus, the caching approach may lack information and provide outdated decisions.

We use APLCache as a running example to explain the details of our proposed approach in the following sections. Thus, the target type of event execution, in this case, are \emph{methods}. APLCache is also used as a baseline for our evaluation, where we investigate the benefits of the two-phase monitoring to APLCache in terms of the overhead and relevance of the provided execution traces. Details about the evaluation are presented in Section~\ref{sec:evaluation}.

\subsection{Coarse-grained Monitoring}
\label{sec:coarse-grained}

In its first phase, our proposed approach monitors the application in a way that it is possible to capture data that enables the identification of relevant traces with low overhead. First, it is necessary to instantiate the solution by providing domain-specific information in terms of relevance criteria and metrics. Based on such information, the coarse-grained phase can collect data to identify the most relevant subset of events of the application. We next detail the manual input required by the coarse-grained phase and how it is used to monitor events.

The coarse-grained phase of the proposed monitoring approach relies on two inputs from the user: (a) the definition of high-level relevance filters using the TigrisDSL language; and (b) the selection of metrics to be used as a quantitative measurement of each relevance criterion referred in filters. To provide such information, users are provided with the guidance derived from our systematic literature review. Based on the user's goal, a set of suitable criteria from those presented in Table~\ref{tab:relevance_criteria} must be selected to be used in relevance filters, and corresponding metrics should be indicated. The metrics presented in Table~\ref{tab:relevance_criteria_metrics} are the most frequently used metrics in the literature, and can be used to represent the desired relevance criteria.

Considering that our running example is focused on identifying suitable method calls for caching, according to Table~\ref{tab:relevance_criteria_metrics}, its goal of monitoring is related to \emph{efficiency}. Thus, the relevance criteria to achieve this goal should include the most popular criteria of this goal group, which are \emph{frequency} and \emph{expensiveness}. In addition, to find caching opportunities, method calls that always provide the same output given a specific input are well-suited for caching due to reuse opportunity~\cite{Mertz2018a,Mertz2020}. Thus, \emph{changeability} is also a criterion considered. As result, we specify the following relevance criteria.

\begin{framed}
\noindent \texttt{(more frequent $\cup$ most expensive) $\cap$ least changeable}
\end{framed}

The TigrisDSL specification presented above contains different modifiers and operators. We are interested in filtering the \emph{more} frequent method calls, i.e.\ the subset of method calls that ranks higher according to a specified metric for frequency, because caching methods that are frequently executed usually leads to performance improvements~\cite{Mertz2020}. In addition to the frequent method calls, we select the \emph{most} expensive method calls, as caching a result that takes more time to be processed would lead to major performance benefits. However, managing cache consistency is a major challenge in the area~\cite{Mertz2020}. Thus, we only include method calls that are the \emph{least} changeable because it would allow us to use a simple consistency strategy, such as an expiration time, and reduce the chances of caching stale content for longer periods. The detailed semantics of the modifiers (\emph{more}, \emph{most}, and \emph{least}) and operators ($\cup$ and $\cap$), as defined by our proposed approach, are presented in Section~\ref{sec:fine-grained}.

For each criterion presented in the TigrisDSL specification, it is necessary to use of a quantitative metric so that the approach is able to track and give an objective interpretation to the criteria. As mentioned, this metric can be any of the metrics presented in Table~\ref{tab:relevance_criteria_metrics}, primarily those that match the selected criteria and the intended goal of monitoring. By taking into account the goal of our running example, possible metrics for frequency, expensiveness and changeability are, respectively, the absolute number of times a method occurs, the average time taken to execute a method and the absolute number of times that each pair of input and output of method occurs.

With a quantitative way of measuring the relevance criteria, the coarse-grained phase starts collecting these metrics from events at runtime. The coarse-grained monitoring results in a summary of the application in terms of statistics about all the event executions of the system. Collected metrics about the execution events are maintained in memory. Consequently, these estimations of the metrics are computationally cheaper than the metrics and do not demand recording individual traces.

These metrics are used as a reference to assess how expensive, frequent, and changeable methods are. For example, considering the computation pattern (changeability), methods with higher standard deviation are less changeable than others because it might indicate that there are equal outputs that are (much) more frequent than others, causing the standard deviation to be high. In our running example, by monitoring events in a coarse-grained manner, our approach gives as result the information presented in Table~\ref{tab:collected_metrics_example}, where a single metric value represents each relevance criterion for each method.

\begin{table*}[!t]
\centering
\caption{Running Example Collected Metrics. Example of metrics collected and maintained for each method in the coarse-grained phase. Cells in gray highlight the \texttt{more frequent}, \texttt{most expensive} and \texttt{least changeable} according to the Grouping step.}
\label{tab:collected_metrics_example}
\begin{tabularx}{\linewidth}{l X X X}
\toprule
\textbf{Method} & \textbf{Frequency} & \textbf{Expensiveness} & \textbf{Changeability} \\ \midrule
ClinicService.findOwner(args) & 12 (less) & 180 (least)  & 6 (more) \\
ClinicService.updateOwner(args) & 2 (least) & \cellcolor{gray!25} 500 (most) & 0 (most) \\
VisitController.newVisit(args) & 50 (frequent) & 250 (more) & 12 (changeable) \\
ClinicService.findVets() & \cellcolor{gray!25} 200 (most) & 300 (expensive) & \cellcolor{gray!25} 200 (least) \\
OwnerRepository.findAll() & \cellcolor{gray!25} 100 (more) & 200 (less) & 90 (less) \\ \bottomrule
\end{tabularx}
\end{table*}

\subsection{Fine-grained Monitoring}
\label{sec:fine-grained}

\begin{figure}[!t]
    \centering
    \includegraphics[width=\linewidth]{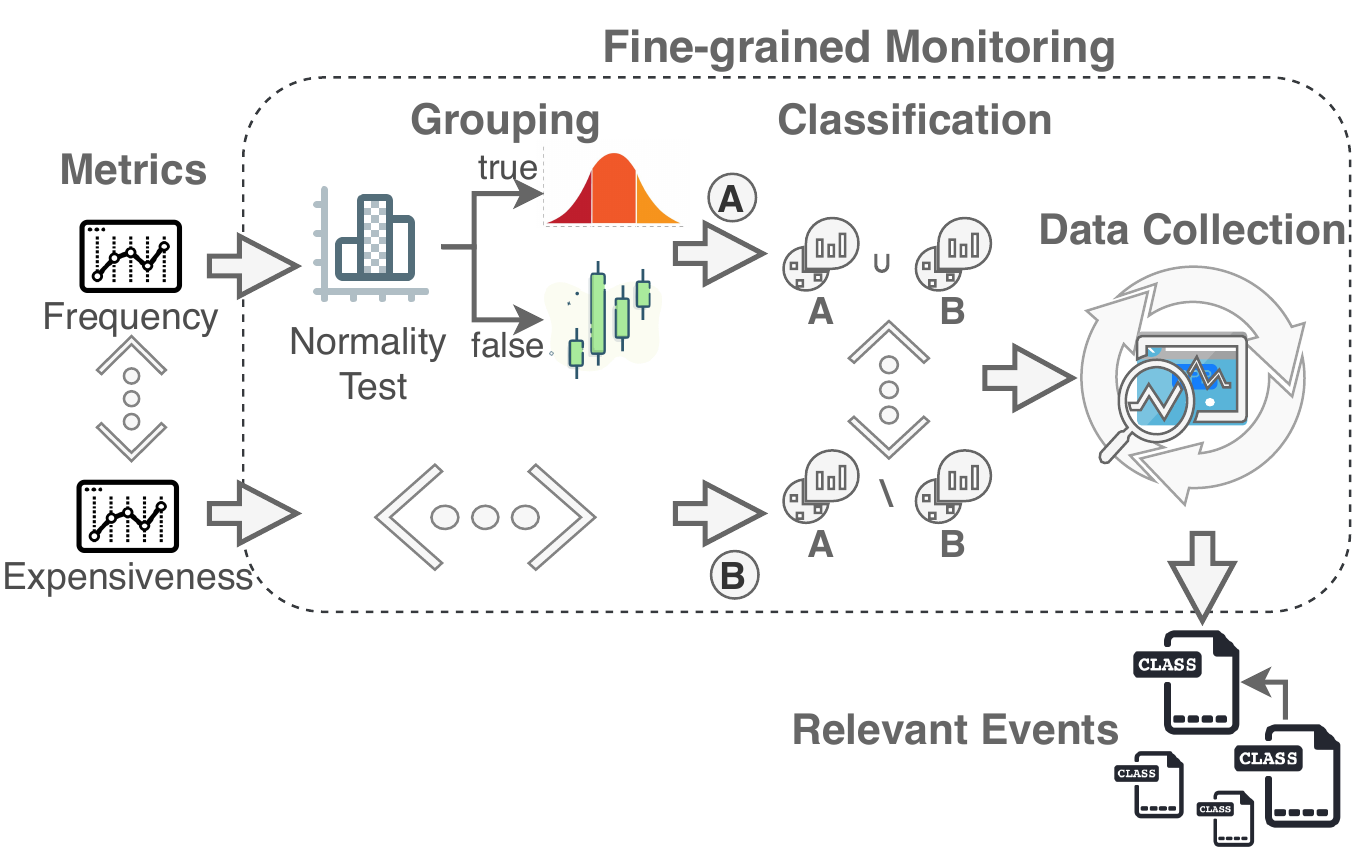}
    \caption{Fine-grained Monitoring Steps. Illustration of the three steps that comprise the fine-grained monitoring: Grouping, Classification and Data Collection.}
    \label{fig:fine_grained}
\end{figure}

By having a summary of the application in terms of statistics about all the event executions of the system and the relevance criteria, it is possible to identify relevant events that should be monitored at runtime in the fine-grained phase. In this phase, our approach uses the data collected in the previous phase to determine which parts of the monitored software system are relevant to the goal of monitoring and thus should be inspected in detail. Figure~\ref{fig:fine_grained} presents an overview of the steps performed during the fine-grained monitoring. The identification of relevant parts happens periodically and consists of processing the calculated metrics to group the values collected in partitions (\emph{Grouping} step) and using such partitions and relevance filters to classify which types of event executions are relevant (\emph{Classification} step). This process generates or updates a list of relevant types of event, which are monitored at a fine-grained level (\emph{Data Collection} step).

\paragraph{Grouping} 

The relevance filters specified using TigrisDSL refer to types of events that satisfy a set of criteria, with each criterion possibly having a modifier (\emph{least}, \emph{less}, \emph{more} and \emph{most}). A criterion with a modifier is used to indicate if events of interest are those that have an associated metric value that is very low, low, average, high, or very high. This way of referring to execution events and their metric values is subjective, and the \emph{Grouping} step uses the distribution (average and spread) of the values of each metric to give an objective meaning.

First, to understand how metric values are distributed, we apply a normality test to the data set (the set of collected metrics for each event execution) to verify whether the observed values of a criterion follow a normal distribution. This process can be performed based on different statistical tests. We adopt the non-parametric statistical test Kolmogorov-Smirnov ($p > 0.05$), which is widely used and performs better with large sample sizes---generally the case when dealing with execution traces. If the Kolmogorov-Smirnov significance value is higher than the alpha value (0.05), then the data follows a normal distribution.

\begin{figure*}[!t]
    \centering
    \begin{subfigure}{0.49\linewidth}
        \includegraphics[width=\linewidth]{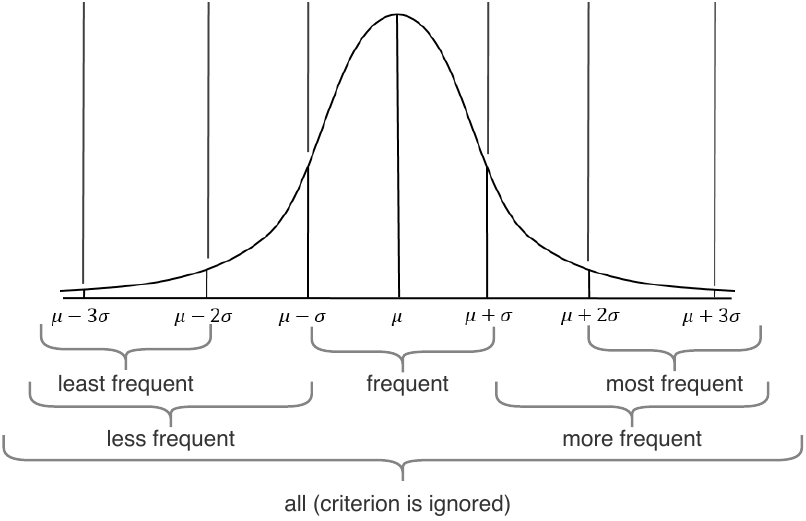}
        \caption{Grouping based on Normally Distributed Data}
        \label{fig:nd}
    \end{subfigure}
    \begin{subfigure}{0.49\linewidth}
        \includegraphics[width=\linewidth]{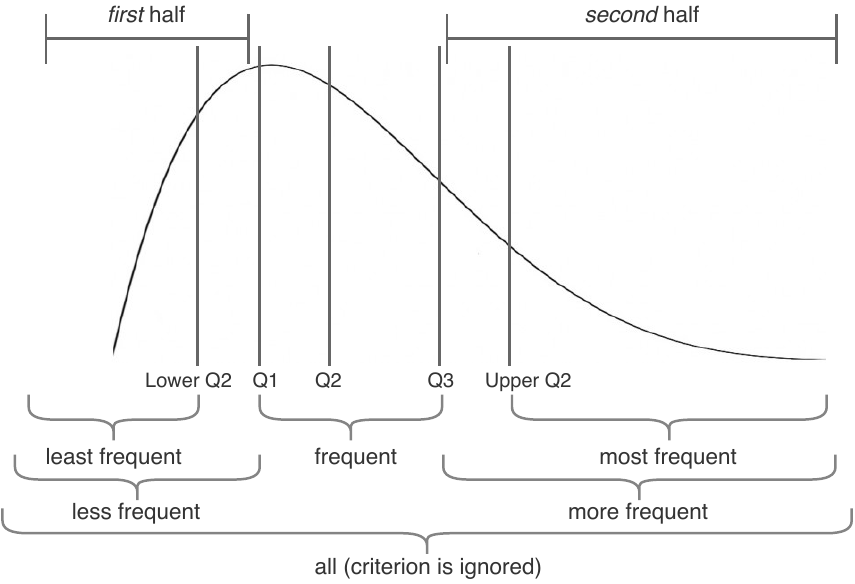}
        \caption{Grouping based on Quantiles}
        \label{fig:quantile}
    \end{subfigure}
    \caption{Frequency Groups. Using the frequency relevance criteria as example, the charts presents the semantics given for the different modifiers of TigrisDSL. The semantics considers the splitting of the data into groups and the normality of the data.}\label{fig:slices}
\end{figure*}

Based on the shape of the distribution, we have two different strategies to classify data into five partitions. In the case of a normal distribution, we group the data based on $K$ standard deviations below or above the mean to create five groups of data. In case the data do not follow a normal distribution, we apply the quantiles strategy by calculating the Q1, Q2, and Q3 of the sample to obtain three groups of data: the lower quarter (below Q1), the interquartile range (between Q1 and Q3) and the upper quarter (above Q3). Then, we calculate the median of the upper and lower quarter and split each of them according to that value, leading to five groups of data. These strategies to group values are presented in Figure~\ref{fig:slices}.

In our running example, we must group methods to identify those that are \texttt{more frequent}, \texttt{most expensive}, and \texttt{least changeable}. The normality test has shown that all distributions are not normal (this is expected as our running example considers a small set of methods). Then, using quantiles to classify the data, we obtain the groups presented in Table~\ref{tab:collected_metrics_example}. In this table, we also highlight the group of interest.

\paragraph{Classification}

The Grouping step evaluates each criterion individually and creates groups of events. The Classification step relies on these groups to evaluate the filters specified in TigrisDSL. This is done by evaluating the provided filter expressions that contain set operations (union, intersection, and subtraction).
The filter specified in our running example is \texttt{(more frequent $\cup$ most expensive) $\cap$ least changeable}. Considering the groups presented in  Table~\ref{tab:collected_metrics_example}, the only method that is relevant considering our goal is \texttt{ClinicService.findVets()}, which satisfies the informed filter. This is the method that should be monitored in detail, in the last step of this phase.

\paragraph{Data Collection}

The list of relevant types of event is updated periodically by the previous steps. It is used in the Data Collection step to perform in-depth monitoring of the relevant event types, obtaining details of their execution, such as returned objects and parameters provided as input. Although filtering a subset of relevant methods reduces the overhead of monitoring, it may still impact the application performance if the traffic is concentrated in those methods supposed to be relevant. Thus, in addition to filtering, traces are collected according to a specified \emph{sampling rate}, which bounds the cost of monitoring. Consequently, our proposal allows the user to achieve an efficient trade-off between sample representativeness and monitoring overhead.
Considering our running example and a sampling rate of 50\%, the method calls that are traced are those highlighted in Table~\ref{tab:data_collection_example}, that is, 50\% of the calls to the \texttt{ClinicService.findVets()} method.

\begin{table}[!t]
\centering
\caption{Sample Execution of an Application. Cells in gray highlight the occurrences (\#Occ.) of a specific method call in the application execution sequence (\#Seq.) that would be traced at a sampling rate of 50\%.\label{tab:data_collection_example}}
\begin{tabular}{llr}
    \toprule
    \textbf{\#Seq.} & \textbf{Method Call} & \textbf{\#Occ.} \\ \midrule
    1 & VisitController.newVisit("X") & 1 \\
    2 & OwnerRepository.findAll() & 1 \\
    3 & ClinicService.findVets() & 1 \\
    4 & \cellcolor{gray!25}ClinicService.findVets() & \cellcolor{gray!25} 2 \\
    5 & ClinicService.findVets() & 3 \\
    6 & ClinicService.updateOwner("X") & 1 \\
    7 & OwnerRepository.findAll() & 2 \\
    8 & \cellcolor{gray!25}ClinicService.findVets() & \cellcolor{gray!25} 4 \\
    9 & VisitController.newVisit("X") & 2 \\
    10 & ClinicService.updateOwner("X") & 2 \\
    11 & ClinicService.findVets() & 5 \\
    12 & \cellcolor{gray!25}ClinicService.findVets() & \cellcolor{gray!25} 6 \\
    13 & ClinicService.findVets() & 7 \\
    14 & ClinicService.updateOwner("X") & 3 \\
    15 & \cellcolor{gray!25}ClinicService.findVets() & \cellcolor{gray!25} 8 \\
    16 & OwnerRepository.findAll() & 3 \\
    ... & ... & ... \\
    $n-1$ & ClinicService.findVets() & $n-1$ \\
    $n$ & \cellcolor{gray!25}ClinicService.findVets() & \cellcolor{gray!25} $n$ \\ \bottomrule
    \end{tabular}
\end{table}

\section{Tigris Framework}
\label{sec:framework-section}

The conceptual approach described above provides a generic means for monitoring software systems and can be instantiated to particular technologies. However, to provide concrete support to this activity and evaluate our approach, it has been implemented as a framework, namely Tigris\footnote{\url{http://prosoft.inf.ufrgs.br/git/Repository/Tree/d32a32bf-1a9e-45e1-8117-b4d2adf3c106}}, using particular technologies. Tigris is implemented in Java and thus can be instantiated and integrated into monitoring-based approaches to leverage the monitoring results of Java-based approaches and applications. This choice is due to our previous programming experience and available tools that were adopted as part of our implementation. To collect data during the coarse-grained and fine-grained monitoring phases, we intercept method executions with AspectJ\footnote{\url{https://eclipse.org/aspectj/}}, which allows Tigris to acquire lightweight and dynamic software metrics without changing the base code. Metrics available in the framework are the most frequently used metrics in the literature, listed in Table~\ref{tab:TigrisMetrics}.

\begin{table*}[!t]
    \caption{Tigris Framework Metrics. List of names and descriptions provided by the Tigris framework, together with how these metrics are estimated.\label{tab:TigrisMetrics}}
        \begin{tabularx}{\linewidth}{l X X}
        \toprule
        \textbf{Metric} & \textbf{Description} & \textbf{Estimation} \\
        \midrule
        Concurrency Level & the number of times a method is executed concurrently & mean number of active threads during all calls of the method \\
        
        Computation Pattern & the number of times that each pair of input and output of method occurs & standard deviation of the return size of all calls of a method  \\
        
        Energy Consumption & the amount of energy demanded by a method & mean estimate of energy consumption of all calls of the method \\
        
        Error level & the number of times the execution of a method thrown exceptions & absolute number of exceptions raised by all calls of the method \\
        
        Execution Time & the time taken to execute a method & mean execution time of all calls of the method \\
        
        Inter-Arrival Time & the time taken between executions of a method & mean time between each call of a method and the next \\
        
        Invocation Frequency & the number of times a method occurs & absolute number of calls of a method \\
        
        Memory Consumption & the amount of memory demanded by a method & mean estimate memory consumption of all calls of the method \\
        
        User Behavior & the number of times a method is shared among different users & absolute number of user sessions that lead to calls of a method \\
        \bottomrule
        \end{tabularx}
\end{table*}

The metrics used to assess relevance criteria might be costly to be collected, because they may require ``heavy'' information to be calculated, such as parameter and return values of event executions. However, as the first phase monitors all system events, this monitoring is coarse grained and collects \emph{lightweight} versions of the metrics to avoid disrupting the application execution. Therefore, although these metrics have some impact on the application execution, it is much lower than collecting fine-grained (and heavy) information and recording execution traces.

The coarse-grained monitoring phase is thus limited to the computation of \emph{estimations} of the metrics---listed in the third column of Table~\ref{tab:TigrisMetrics}---which are kept as a single in-memory number, continuously updated whenever a new event is intercepted at runtime. For example, for execution time and invocation frequency, the estimations are the mean execution time and the absolute number of all the calls of a method, respectively. This is opposed to the complete distribution of these metrics with traces giving information such as which inputs lead to high execution times.

Computation pattern, in turn, consists of the analysis of common computations of a method, that is, the identification of frequent pairs of inputs and output of a method. This is expensive to be computed as it requires tracing and comparing each method call accompanied by the parameter values and the method return value. Because the goal of this metric is to identify repeated computations and the output of a method is usually highly dependent on its input, our estimation relies on observing the return values of different calls of a method in terms of allocation size in memory and then computing the standard deviation of these values. Thus, for example, if the standard deviation is low, it means that most of the returns of the method calls are the same, thus less changing.

For estimations that would demand to store the list of observed values such as those based on standard deviation and average, to keep a single in-memory number updated on-the-fly, we compute mean and standard deviation based on online and incremental algorithms~\cite{Knuth1997}. The TigrisDSL specification, as well as metrics to be used while assessing execution events can be configured through property files and annotations. An example of such configuration is presented in Listing~\ref{lst:configExample}.

\begin{lstlisting}[caption={Tigris annotation-based configuration example.},label={lst:configExample},float]
@TigrisConfiguration(
    logRepository = RepositoryType.MEMORY,
    staticMetricFile = "petclinic.csv",
    samplingPercentage = 0.5,
    analysisFixedDelay = 120)
@TigrisCriteria(
    criteria = "more frequent U more expensive",
    granularity = GranularityType.METHOD,
    frequencyMetric = Metrics.INVOCATION_FREQUENCY,
    expensivenessMetric = Metrics.EXECUTION_TIME,
    changeabilityMetric = Metrics.COMPUTATION_PATTERN)
@ComponentScan(allowed = "org.springframework.samples.petclinic.*", 
    denied = "org.springframework.samples.petclinic.model.*")
public class Configuration {...}
\end{lstlisting}

The sampling rate can be controlled and adjusted at runtime (using a function provided by our implementation), varying from 0\% (no monitoring) to 100\% (complete monitoring of selected methods). This function can be used to adapt the sampling according to the overhead tolerance and monitoring coverage requirement. In addition, the amount of time in which the framework should keep collecting lightweight metrics (first monitoring phase) of the event executions until triggering the process to select methods to monitor in detail (second monitoring phase) can be controlled and adjusted at runtime through an input parameter.

New criteria and metrics can be included by extending and implementing interfaces provided by the framework. The same interfaces are used to customize and define how metrics should be calculated. In addition to the usage of TigrisDSL filters to identify relevant event executions based on the goal of monitoring, Tigris also offers customizations. For example, it is possible to set up the framework to focus on specific monitoring locations, which can improve the set of events to be evaluated as relevant as well as exclude events that must not be monitored, thus reducing the time overhead for tracking them. Tigris also supports loading output metrics from Understand\footnote{\url{https://scitools.com/}}, to evaluate criteria that are based on static metrics.

Next, we describe our evaluation procedure and then discuss the obtained results.

\section{Evaluation: Adaptive Monitoring for APLCache}
\label{sec:evaluation}

In order to evaluate our proposed solution for monitoring, we perform an empirical evaluation by instantiating Tigris as monitoring support for APLCache, the application-level caching approach that was used to illustrate the phases of our framework.

\subsection{Study Settings}\label{subsec:study_settings}

APLCache~\cite{Mertz2018a} is used to monitor web applications to identify cacheable methods with the goal of improving application performance. The monitoring of APLCache captures execution traces of each method call with its input parameters and return. Monitoring the application has a performance cost, so we aim with Tigris to reduce the monitoring cost without compromising its effectiveness of identifying cacheable methods. Therefore, the original version of APLCache is used as a baseline.

Our evaluation aims to answer three research questions, presented in Table~\ref{tab:rqs}. In this table, we also detail the metrics used to answer each research question. In RQ1, we assess how Tigris reduces the cost of monitoring. However, as the application performance is influenced not only by the monitoring but also by the identified cacheable spots that are discovered based on the collected traces, we also compare the performance of our target web applications using APLCache with full monitoring and with Tigris. In RQ2, we compare the effectiveness of the cacheable spots identified using the traces collected using the two alternatives under evaluation. Improved effectiveness to identify cacheable methods is desirable. Nevertheless, our ultimate goal is to monitor the application with lower costs, without compromising the results of the analysis, that is, we aim to collect a subset of execution traces that would lead to the same results as if we had collected the complete set of traces. Thus, we also assess the effectiveness (precision and recall) of APLCache with Tigris using as ground-truth the cacheable spots identified by APLCache with full monitoring. Finally, in RQ3 we assess how Tigris copes with workload variations over time in terms of changes in the relevance evaluation performed by its first phase (coarse-grained monitoring) and, consequently, in the selection of execution traces to be collected in detail by its second phase (fine-grained monitoring).

\begin{table*}[!t]
    \caption{Research Questions and Metrics. List of the research questions of our evaluation and the metrics used to answer each of them.}
    \label{tab:rqs}
    \begin{tabularx}{\linewidth}{p{5.5cm} X}
    \toprule
    \textbf{Research Question} & \textbf{Metric} \\ \midrule
    \textbf{RQ1.} What performance gain does Tigris provide? & 
        \textbf{M1-1.} Throughput (average number of requests handled per second) of the target applications with monitoring (and no caching)
        \par \textbf{M1-2.} Throughput of the target applications using APLCache
    \\ \midrule
    \textbf{RQ2.} What is the effectiveness achieved with execution traces collected by Tigris? &
        \textbf{M2-1.} Number of identified caching opportunities
        \par \textbf{M2-2.} Hit ratio
        \par \textbf{M2-3.} Number of hits
        \par \textbf{M2-4.} Precision
        \par \textbf{M2-5.} Recall
        \par \textbf{M2-6.} F-measure
    \\ \midrule
    \textbf{RQ3.} How does Tigris cope with workload variations over time? & 
        \textbf{M3-1.} Difference in the number methods selected by the first phase of the approach (coarse-grained monitoring) through multiple monitoring cycles in sequence
    \\ \bottomrule
\end{tabularx}
\end{table*}

To assess both versions of APLCache, we must select target web applications, simulation parameters, and workloads. To avoid bias, we follow the design choices of the study previously performed to evaluate the original version of APLCache~\cite{Mertz2018a}. In the study, we use three target open-source web applications\footnote{Available at \url{http://www.cloudscale-project.eu/}, \url{https://github.com/SpringSource/spring-petclinic/} and \url{http://www.shopizer.com/}.}, presented in Table~\ref{tab:targetsystems}, which summarizes the general characteristics of each target system. We also keep the same APLCache parameters, such as cache provider and eviction policy.

For all the RQs we consider performance test suites in the form of workload simulations. The simulation starts with five simultaneous users continuously navigating through the application based on a navigation pattern that falls into a specific distribution (transition table). Then, at every second, we randomly add or remove a number of users (from 1 to 10) from the simulation until all the users make the total of 60k requests to the application. We adopt a minimum number of 5 simultaneous users to keep a minimum of concurrency in the workload, and a maximum of 20 to avoid disruptions in the response times due to struggles from the web server. To stimulate changes in the workload, we created three variations of navigation patterns for each application, and whenever a user is added to the simulation, we randomly decide which of the three workload variations the new user should follow. To keep a fresh selection of execution traces collected in detail by the second phase (fine-grained monitoring), the processing of lightweight metrics is triggered every two minutes during the simulation.

To increase the reliability of the results, we create the workload with the above settings once and execute the exactly same sequence of requests and user variations per second ten times. To evaluate how changes in the workload may impact in our proposal, the simulation is segmented in three \emph{monitoring cycles} of 20k requests, we collect and inspect from the simulation the subset of methods selected by the first phase of Tigris as well as the cacheable opportunities found by APLCache. Thus, for all the metrics in the results we report mean and standard deviations of these multiple executions. For all the simulations, we use two machines located within the same network, one machine (16G RAM, Intel i7 2GHz) for the Tomcat\footnote{\url{http://tomcat.apache.org/}} web server and MySQL\footnote{\url{https://www.mysql.com/}} database, and one machine (16G RAM, Intel i5 2.4GHz) to handle the performance test suite with JMeter\footnote{\url{http://jmeter.apache.org/}}.

\begin{table}[!t]
    \caption{Target Systems. List of the target web applications used in our evaluation, together with their application domain and size. Size is detailed by the number of lines of code (LOC) and the number of files.}
    \label{tab:targetsystems}
		\setlength{\tabcolsep}{1mm}
    \centering
    \begin{tabular}{l l rr}
      \toprule
      \textbf{Project} & \textbf{Domain} & \textbf{LOC} & \textbf{\# Files} \\
      \midrule
      Cloud Store & File Synchronization & 7.6 K & 98 \\ 
      Pet Clinic & Sample application & 6.3 K & 72 \\ 
      Shopizer & e-Commerce & 111.3 K & 946 \\ 
      \bottomrule
    \end{tabular}
\end{table}

To configure Tigris, we must specify relevance filters in TigrisDSL. We used as a basis the Cacheability Pattern~\cite{Mertz2017a}, which indicates a set of criteria for deciding whether a method should be cached. We assess two alternative filters in our study, presented in Table~\ref{tab:relevance_input}. The \emph{Restricted Filter} leads to a subset of methods selected by the \emph{Expanded Filter} to be monitored in the second phase of our framework. The specified relevance filters indicate that four relevance criteria are considered and, for each of them, we must indicate the metric estimations to be used. We selected the metrics Invocation Frequency, Computation Pattern, User Behavior and Execution Time, for the criteria Frequency, Changeability, Usage pattern and Expensiveness, respectively. The Tigris second phase also receives as parameter a sampling rate. We selected six sampling rates (ranging from 1\% to 100\%) to understand how the number of collected traces can impact in the analysis of the traces.

\begin{table}[!t]
\caption{Relevance Filter Specification. Specification of the two filters used in our evaluation, namely Restricted Filter and Expanded Filter.}
\label{tab:relevance_input}
\begin{tabularx}{\linewidth}{p{1.5cm} X}
\toprule
\textbf{Name} & \textbf{TigrisDSL-based Specification} \\ \midrule
 Restricted Filter & \texttt{less changeable $\cap$ more frequent $\cap$ (more usage-pattern $\cup$ (more expensive $\cap$ less usage-pattern))} \\ \midrule
 Expanded Filter & \texttt{(less changeable $\cup$ changeable) $\cap$ (more frequent $\cup$ frequent) $\cap$ ((more usage-pattern $\cup$ usage-pattern) $\cup$ (more expensive $\cup$ expensive))} \\ \bottomrule
\end{tabularx}
\end{table}

\subsection{Results}

We next present and analyze the results obtained by running the simulations with each of our three target applications and collecting the specified metrics.

\subsubsection{RQ1. What performance gain does Tigris provide?}

Our first results consist of assessing how much Tigris reduces the cost of monitoring, considering its ability to filter and reduce the scope of monitoring. For that, we compare the performance of each application using Tigris (restricted and expanded filters) to the baselines with full monitoring and no monitoring. The information collected with full monitoring leads to the ground-truth decisions made based on execution traces, while no monitoring provides a baseline of the application performance without any overhead. This analysis allows us to assess how far our decisions are from the ground truth and the performance costs associated with them. To compare the application performance under these different monitoring configurations, we use \emph{throughput}, which is measured by calculating the average number of requests handled per second throughout the simulation. Thus, high throughput (i.e.\ close to the throughput achieved with no monitoring) indicates an effective monitoring configuration, as more requests can be processed within the same period of time.

\begin{figure}[!t]
    \centering
    \begin{subfigure}{\linewidth}
        \includegraphics[width=\linewidth]{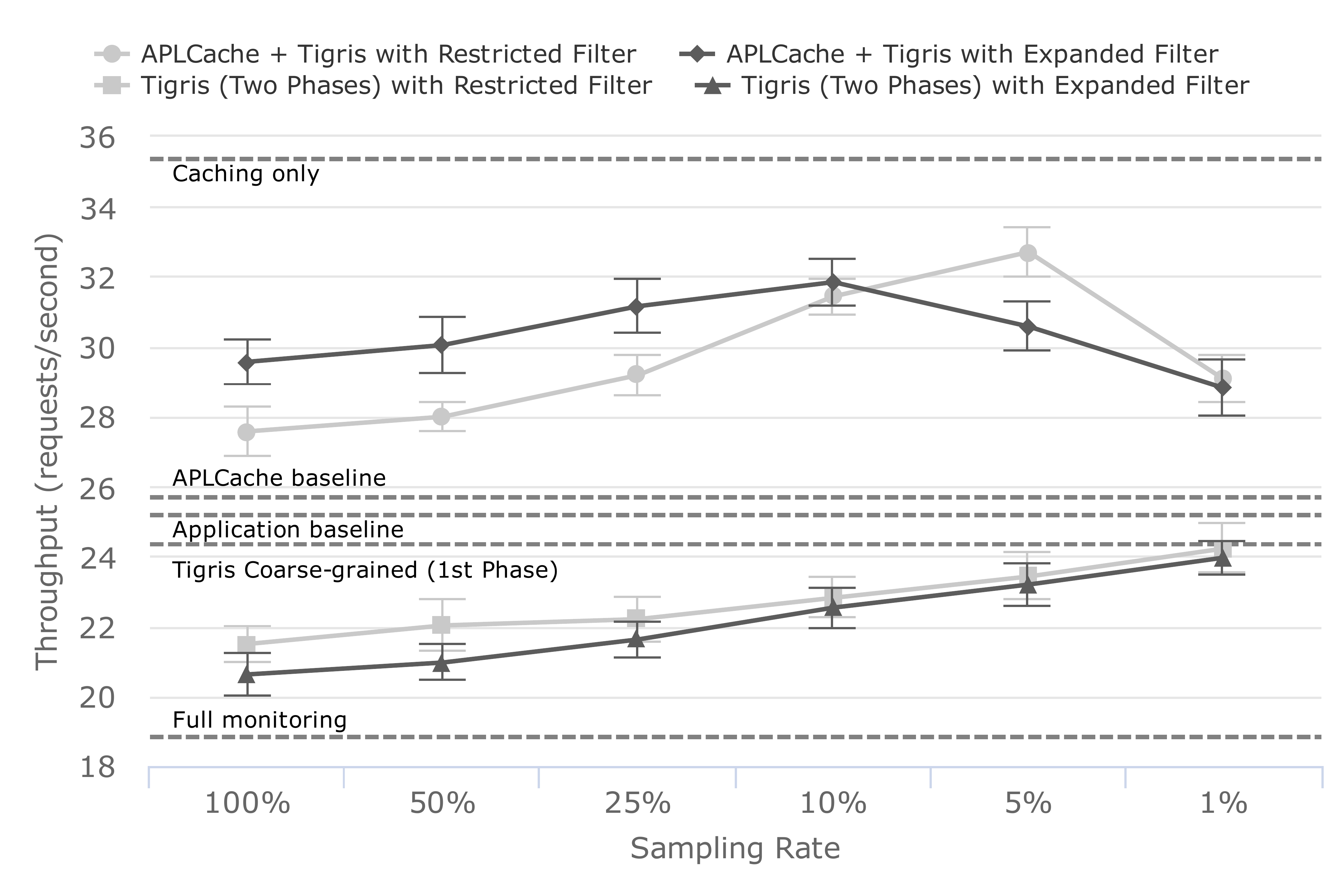}
        \caption{Cloud Store}
        \label{fig:cloudstore}
    \end{subfigure}
    \begin{subfigure}{\linewidth}
        \includegraphics[width=\linewidth]{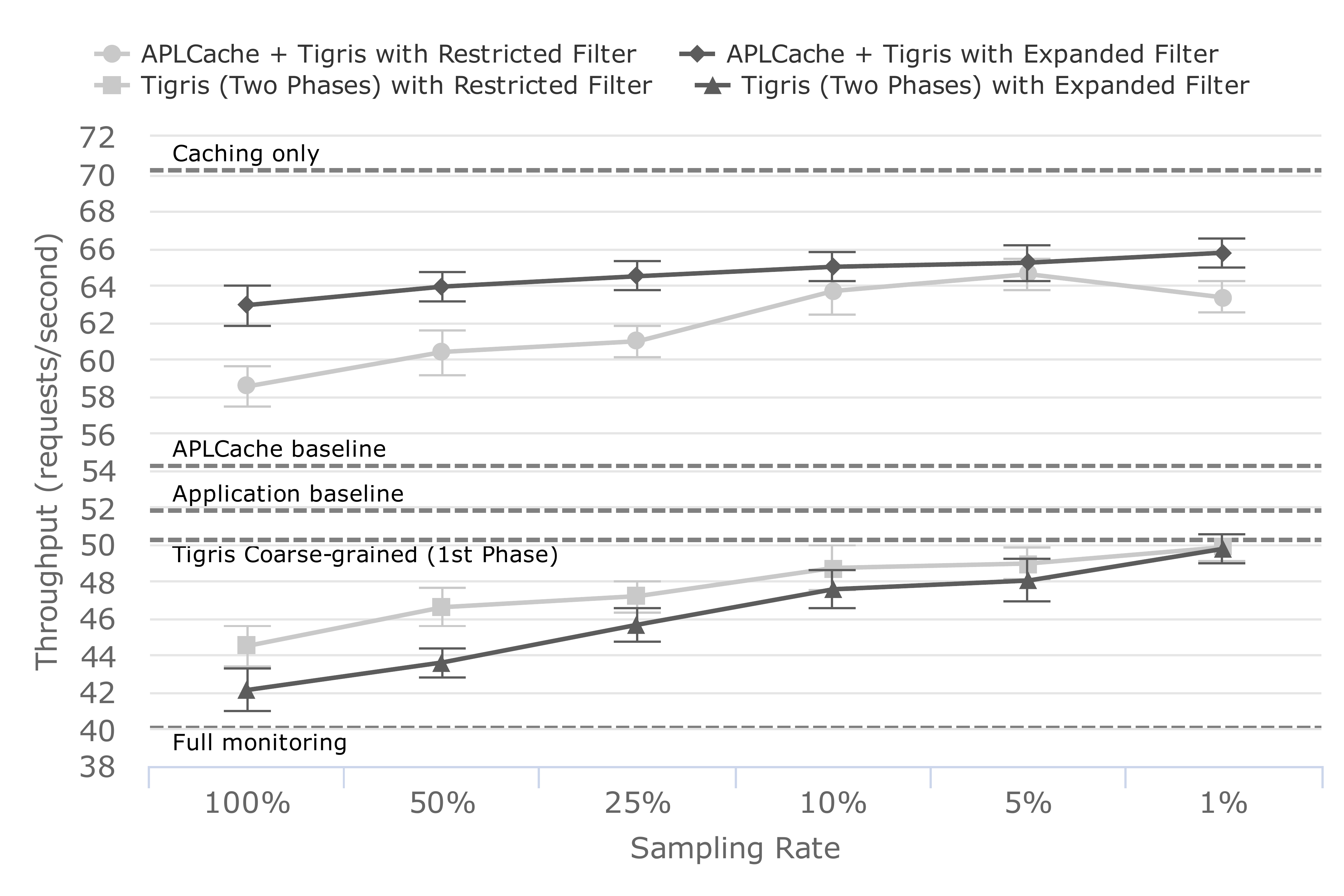}
        \caption{Pet Clinic}
        \label{fig:petclinic}
    \end{subfigure}
    \begin{subfigure}{\linewidth}
        \includegraphics[width=\linewidth]{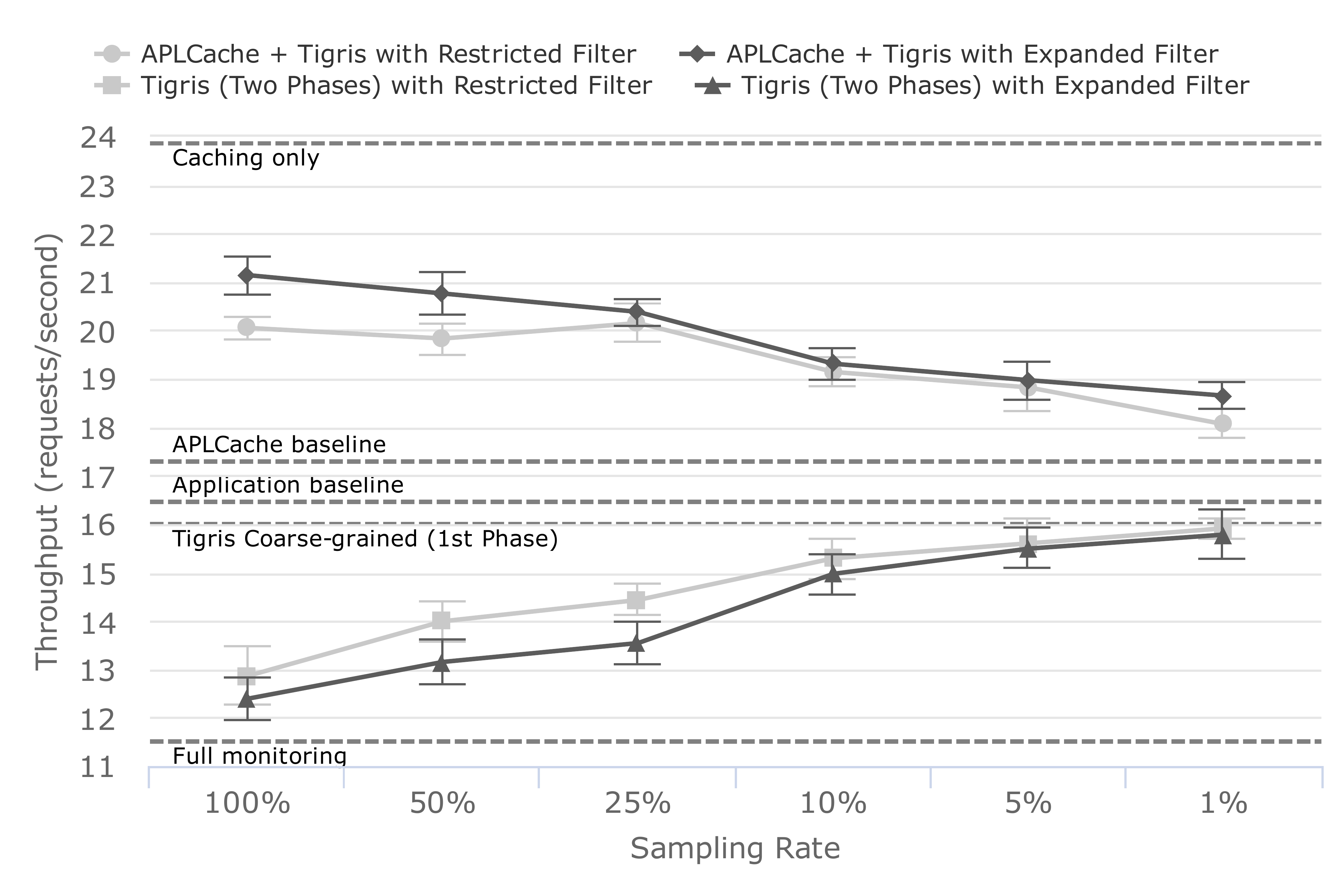}
        \caption{Shopizer}
        \label{fig:shopizer}
    \end{subfigure}
    \caption{RQ1: Throughput by Sampling Rate. Performance of each application executed with and without APLCache using varying monitoring configuration.\label{fig:resultsRQ1}}
\end{figure}

The results are presented in Figure~\ref{fig:resultsRQ1}. In each chart, we present the throughput of executing the application with no and full monitoring, which serve as references. We also detail the cost of running the application only with the Tigris first phase activated. All these are presented as horizontal lines because they do not vary in terms of the sampling rate. The throughput of running Tigris with the two considered filters is presented with its varying results according to the sampling rate.

As expected, monitoring an application causes an overhead, even with Tigris. Considering the different applications and varying evaluated configurations, we observed that the minimum overhead was obtained with the restricted filter and 1\% sampling rate for Shopizer (Figure~\ref{fig:shopizer}). This configuration achieves a throughput of 15.91 req/s vs.\ 16.47 req/s obtained with no monitoring, resulting in 3.42\% of performance impact. The maximum observed overhead came from the combination of expanded filter and 100\% sampling rate for Shopizer (Figure~\ref{fig:shopizer}), which led to a throughput of 12.39 req/s---24.7\% lower than the baseline with no monitoring (16.47 req/s). In brief, Tigris resulted in a performance penalty ranging from 3.4\% to 24.7\%, compared to the baseline without monitoring.

However, the framework largely reduces the cost of full monitoring. The overhead of this configuration ranges from 22.6\% (its throughput is 40.09 req/s for Petclinic as seen in Figure~\ref{fig:petclinic}, vs.\ 51.81 req/s of the baseline) to 29.9\% (its throughput is 11.53 req/s for Shopizer as seen in Figure~\ref{fig:shopizer}, vs.\ 16.47 req/s of the baseline).

Assessing solely the cost of the Tigris, we observe that the overhead is caused mostly by its second phase, in which methods are fine-grained monitored. The maximum overhead of the coarse-grained monitoring (first-phase) was 3.3\%, because Cloud Store achieved 24.32 req/s when compared to the baseline without monitoring of 25.16 req/s.

The performance overhead of the second phase varies according to the results of the first phase and the sampling rate in which it collects traces. The overhead decreases when the sampling rate decreases because fewer executions are traced. For example, when collecting 1\% of the traces with the restricted filter, the overhead is marginal for all three applications, ranging from 3.4\% to 3.8\% when comparing to the baseline with no monitoring in Figure~\ref{fig:resultsRQ1}, being similar to the coarse-grained phase only. Collecting traces with the expanded filter at a 100\% sampling rate, i.e.\ no sampling is performed, largely increases the overhead to the application performance, ranging from 18.0\% to 24.7\%, when comparing to the baseline with no monitoring in Figure~\ref{fig:resultsRQ1}.

To compare the overall performance of APLCache with full monitoring and with Tigris, we also assess the throughput. However, the overall performance is not only affected by the monitoring but also by the opportunities cached based on the analysis of collected traces. Obtained results are presented in Table~\ref{tab:results} (column Throughput) and Figure~\ref{fig:resultsRQ1}. We observe that the throughput achieved by APLCache when supported by Tigris is higher than using full monitoring, with both filters and any sampling rate. Tigris improves the throughput of all the three target applications with gains ranging from 4.4\% to 27.4\%, in comparison to APLCache with full monitoring. Because Tigris can filter a subset of relevant methods and consequently monitor fewer methods in detail, the overhead of monitoring tends to decrease.


\begin{table*}[!t]
\centering
\caption{Simulation Results. Results of executing each application with full monitoring, restricted filter and expanded filter. Reported metrics (average of all the ten executions) for the different sampling rates (Sample) are throughput, hit ratio (HR), number of hits (Hits), number of cacheable opportunities (Cacheability), precision (Pr.), recall, and F-Measure (F1). Throughput and Cacheability are shown in absolute and relative (percentage change in comparison with full monitoring) terms. Cacheability, precision, recall and F-Measure are presented as the average of all three monitoring cycles.}

\label{tab:results}
\begin{tabular}{llrrrrrrrrrr}
\toprule
& \textbf{Monitoring} & \textbf{Sample} & \multicolumn{2}{c}{\textbf{Throughput}} & \textbf{HR} & \textbf{Hits} & \multicolumn{2}{c}{\textbf{Cacheability}} & \textbf{Pr.} & \textbf{Recall} & \textbf{F1} \\ \midrule \midrule

\multicolumn{1}{l}{\multirow{13}{*}{\rotatebox[origin=c]{90}{\textbf{Cloud Store}}}} 
 & Full Monitoring &  & 25.6 & & 0.96 & 45,877 & 7.67 & \\ \cmidrule(l){2-12}
 
 & \multirow{6}{*}{Restricted Filter} & 100\% & 27.5 & +7.3\% & 0.96 & 30,252 & 3.67 & -52.2\% & 1.0 & 0.47$\pm$0.04 & 0.64$\pm$0.03 \\
                                    &  & 50\% & 27.9 & +9.0\% & 0.95 & 27,036 & 2.67 & -65.2\% & 1.0 & 0.34$\pm$0.05 & 0.51$\pm$0.05 \\
                                    &  & 25\% & 29.1 & +13.7\% & 0.97 & 25,924 & 2.33 & -69.6\% & 1.0 & 0.30$\pm$0.06 & 0.46$\pm$0.07 \\
                                    &  & 10\% & 31.4 & +22.4\% & 0.98 & 26,100 & 2.33 & -69.6\% & 1.0 & 0.30$\pm$0.06 & 0.46$\pm$0.07 \\
                                    &  & 5\% & 32.7 & +27.4\% & 0.97 & 25,532 & 2.33 & -69.6\% & 1.0 & 0.30$\pm$0.06 & 0.46$\pm$0.07 \\
                                    &  & 1\% & 29.0 & +13.2\% & 0.98 & 17,208 & 1 & -87.0\% & 1.0 & 0.13$\pm$0.01 & 0.23$\pm$0.01 \\ \cmidrule(l){2-12}
 
 & \multirow{6}{*}{Expanded Filter} & 100\% & 29.5 & +15.1\% & 0.96 & 42,144 & 6.33 & -17.4\% & 1.0 & 0.83$\pm$0.14 & 0.90$\pm$0.08 \\
                                 &  & 50\% & 30.0 & +17.0\% & 0.94 & 41,668 & 6.33 & -30.4\% & 1.0 & 0.70$\pm$0.18 & 0.81$\pm$0.13 \\
                                 &  & 25\% & 31.1 & +21.3\% & 0.98 & 36,060 & 4 & -47.8\% & 1.0 & 0.52$\pm$0.04 & 0.68$\pm$0.03 \\
                                 &  & 10\% & 31.8 & +24.0\% & 0.94 & 33,384 & 3.67 & -52.2\% & 1.0 & 0.48$\pm$0.09 & 0.64$\pm$0.09 \\
                                 &  & 5\% & 30.5 & +19.1\% & 0.96 & 25,524 & 2.33 & -69.6\% & 1.0 & 0.30$\pm$0.10 & 0.46$\pm$0.11 \\
                                 &  & 1\% & 28.81 & +12.2\% & 0.95 & 20,328 & 1.33 & -82.6\% & 1.0 & 0.17$\pm$0.09 & 0.29$\pm$0.12 \\ \midrule

\multicolumn{1}{l}{\multirow{13}{*}{\rotatebox[origin=c]{90}{\textbf{Petclinic}}}}
 & Full Monitoring &  & 54.17 & & 0.94 & 52,860 & 4 & \\ \cmidrule(l){2-12}
 & \multirow{6}{*}{Restricted Filter} & 100\% & 58.5 & +8.0\% & 0.94 & 41,040 & 2.33 & -41.7\% & 1.0 & 0.58$\pm$0.14 & 0.73$\pm$0.11 \\    
                                    &  & 50\% & 60.3 & +11.4\% & 0.94 & 41,404 & 2.33 & -41.7\% & 1.0 & 0.58$\pm$0.14 & 0.73$\pm$0.11 \\
                                    &  & 25\% & 60.9 & +12.5\% & 0.94 & 38,076 & 2 & -50.0\% & 1.0 & 0.50$\pm$0.00 & 0.66$\pm$0.00 \\
                                    &  & 10\% & 63.6 & +17.5\% & 0.94 & 38,488 & 2 & -50.0\% & 1.0 & 0.50$\pm$0.00 & 0.66$\pm$0.00 \\
                                    &  & 5\% & 64.5 & +19.2\% & 0.94 & 38,980 & 2 & -50.0\% & 1.0 & 0.50$\pm$0.00 & 0.66$\pm$0.00 \\
                                    &  & 1\% & 63.3 & +16.8\% & 0.94 & 26,224 & 1 & -75.0\% & 1.0 & 0.25$\pm$0.00 & 0.40$\pm$0.00 \\ \cmidrule(l){2-12}
 & \multirow{6}{*}{Expanded Filter} & 100\% & 62.9 & +16.1\% & 0.94 & 52,616 & 4 & 0.0\% & 1.0 & 1.00$\pm$0.00 & 1.00$\pm$0.00 \\
                                    &  & 50\% & 63.9 & +17.9\% & 0.94 & 52,772 & 4 & 0.0\% & 1.0 & 1.00$\pm$0.00 & 1.00$\pm$0.00 \\
                                    &  & 25\% & 64.4 & +18.9\% & 0.94 & 49,032 & 3.67 & -8.3\% & 1.0 & 0.91$\pm$0.14 & 0.95$\pm$0.08 \\
                                    &  & 10\% & 64.9 & +19.9\% & 0.94 & 46,900 & 3 & -25.0\% & 1.0 & 0.75$\pm$0.00 & 0.85$\pm$0.00 \\
                                    &  & 5\% & 65.1 & +20.3\% & 0.94 & 46,652 & 3 & -25.0\% & 1.0 & 0.75$\pm$0.00 & 0.85$\pm$0.00 \\
                                    &  & 1\% & 65.7 & +21.3\% & 0.94 & 45,824 & 3 & -25.0\% & 1.0 & 0.75$\pm$0.00 & 0.85$\pm$0.00 \\ \midrule

\multicolumn{1}{l}{\multirow{13}{*}{\rotatebox[origin=c]{90}{\textbf{Shopizer}}}}
 & Full Monitoring &  & 17.30 & & 0.92 & 691,352 & 24.33 & \\ \cmidrule(l){2-12} 
 & \multirow{6}{*}{Restricted Filter} & 100\% & 20.0 & +15.9\% & 0.92 & 419,656 & 16.33 & -32.9\% & 1.0 & 0.66$\pm$0.03 & 0.80$\pm$0.02 \\
                                      &  & 50\% & 19.8 & +14.6\% & 0.93 & 257,012 & 13.33 & -45.2\% & 1.0 & 0.54$\pm$0.01 & 0.70$\pm$0.01 \\
                                      &  & 25\% & 20.1 & +16.4\% & 0.92 & 182,520 & 10 & -59.0\% & 1.0 & 0.40$\pm$0.03 & 0.58$\pm$0.03 \\
                                      &  & 10\% & 19.1 & +10.6\% & 0.94 & 71,460 & 6 & -75.3\% & 1.0 & 0.24$\pm$0.01 & 0.39$\pm$0.02 \\
                                      &  & 5\% & 18.8 & +8.7\% & 0.93 & 35,892 & 4 & -83.6\% & 1.0 & 0.16$\pm$0.02 & 0.27$\pm$0.03 \\
                                      &  & 1\% & 18.0 & +4.4\% & 0.91 & 17,998 & 1 & -95.9\% & 1.0 & 0.04$\pm$0.00 & 0.07$\pm$0.01 \\ \cmidrule(l){2-12} 
                                      
 & \multirow{6}{*}{Expanded Filter} & 100\% & 21.1 & +22.2\% & 0.91 & 513,960 & 17.67 & -27.4\% & 1.0 & 0.74$\pm$0.17 & 0.84$\pm$0.11 \\
                                    &  & 50\% & 20.7 & +20.0\% & 0.91 & 350,072 & 14.33 & -41.1\% & 1.0 & 0.60$\pm$0.14 & 0.74$\pm$0.11 \\
                                    &  & 25\% & 20.3 & +17.8\% & 0.92 & 184,232 & 10.33 & -57.5\% & 1.0 & 0.43$\pm$0.10 & 0.60$\pm$0.10 \\
                                    &  & 10\% & 19.3 & +11.7\% & 0.91 & 68,660 & 6 & -75.3\% & 1.0 & 0.25$\pm$0.06 & 0.40$\pm$0.08  \\
                                    &  & 5\% & 18.9 & +9.7\% & 0.91 & 32,818 & 3.33 & -86.3\% & 1.0 & 0.14$\pm$0.03 & 0.24$\pm$0.05 \\
                                    &  & 1\% & 18.6 & +7.8\% & 0.93 & 18,412 & 1 & -95.9\% & 1.0 & 0.04$\pm$0.00 & 0.08$\pm$0.01 \\ \bottomrule
\end{tabular}
\end{table*}

In Pet Clinic (Figure~\ref{fig:petclinic}), the overall performance typically increases as the sampling rate decreases. Nevertheless, this does not always hold. With fewer execution traces (lower sampling rate), the identification of cacheable opportunities is less consistent with the actual behavior of the application, thus reducing the gains of caching the application. This can be seen in the other two applications. The overall performance of CloudStore (Figure~\ref{fig:cloudstore}) increases up to 5\% sampling rate for the restricted filter and 10\% sampling rate for the expanded filter, and then it decays. The overall performance of the Shopizer application (Figure~\ref{fig:shopizer}), in turn, tends only to decrease as the sampling rate decreases.
In most cases, the restricted filter achieves worse results than the expanded filter. This means that, although the restricted filter leads to fewer methods to be monitored in the second monitoring phase, the set of fine-grained monitored methods causes the identification of caching opportunities that provide lower increases in the performance. That is, the balance between the cost of monitoring and the gain of caching is higher with the expanded filter than with the restricted filter.

\begin{framed}
\noindent \textbf{RQ1: Findings.} The impact of full monitoring applications is high, causing performance impacts ranging from 22.6\% to 29.9\% when compared to the baseline of no monitoring. Tigris, with varying configurations, implies a lower impact to the application performance, with values ranging between 2.9\% and 24.7\%, when compared to the baseline of no monitoring. The overhead of Tigris is mostly caused by fine-grained monitoring, which can be reduced by decreasing the sampling rate. With respect to overall performance with enabled caching, it provides improvements ranging from 4.4\% to 27.4\% in relation to monitoring all method calls with APLCache. The relevance filter and sampling rate provide a configuration space that allows the approach to achieve the best trade-off between the cost of monitoring and the quality of the analysis of execution traces.
\end{framed}

\subsubsection{RQ2. What is the effectiveness achieved with execution traces collected by Tigris?}

The previous research question has shown that the performance gains depend on which methods are cached. This decision is made by APLCache, which analyzes the collected execution traces. Therefore, we now evaluate the cached opportunities identified with each monitoring configuration. We assess the number of cached opportunities, the hit ratio, and the number of hits. These are presented in Table~\ref{tab:results} (columns Cacheability, HR and Hits, respectively).

\begin{figure}[!t]
    \centering
    \begin{subfigure}{\linewidth}
        \includegraphics[width=\linewidth]{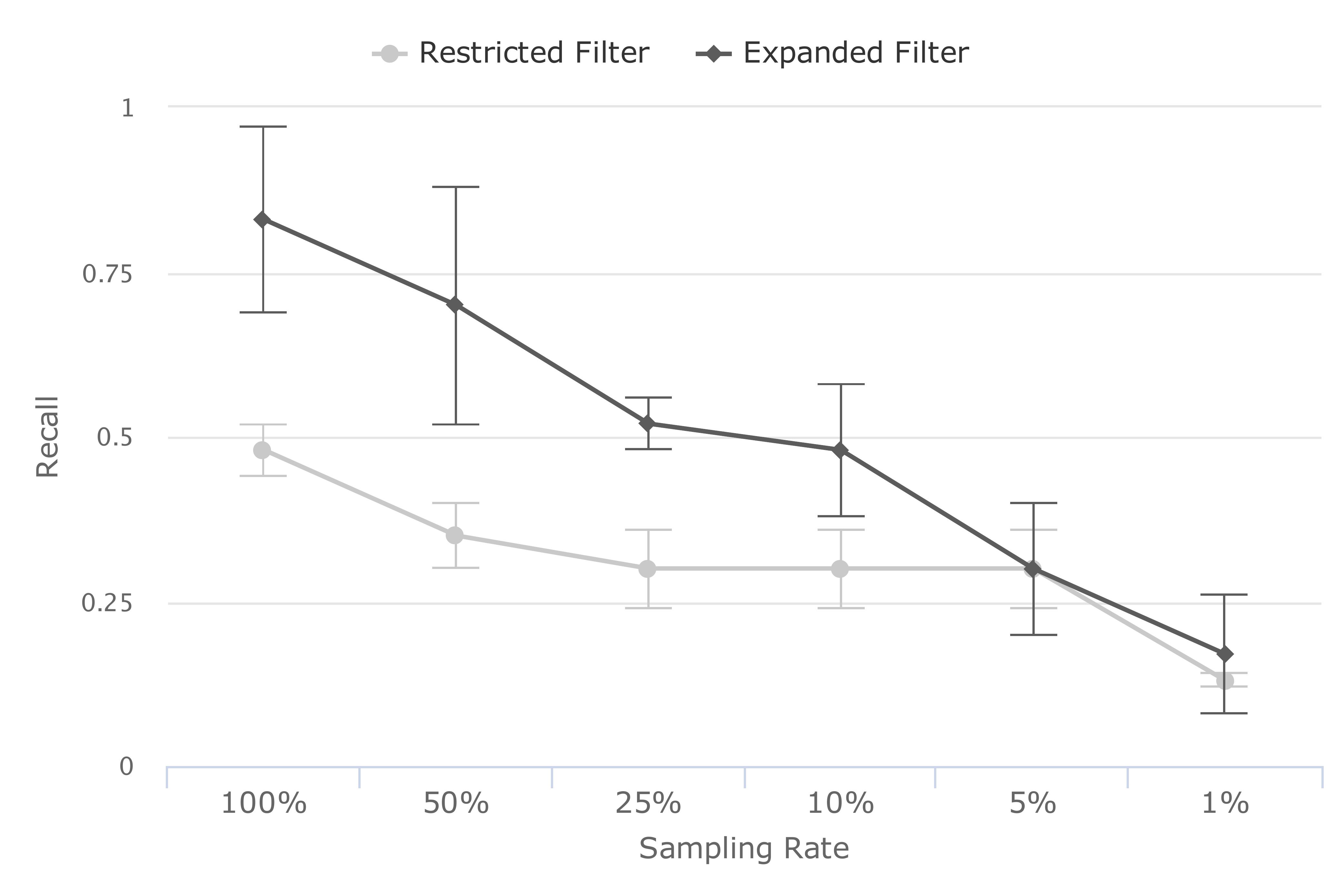}
        \caption{Cloud Store}
        \label{fig:recall-cloudstore}
    \end{subfigure}
    \begin{subfigure}{\linewidth}
        \includegraphics[width=\linewidth]{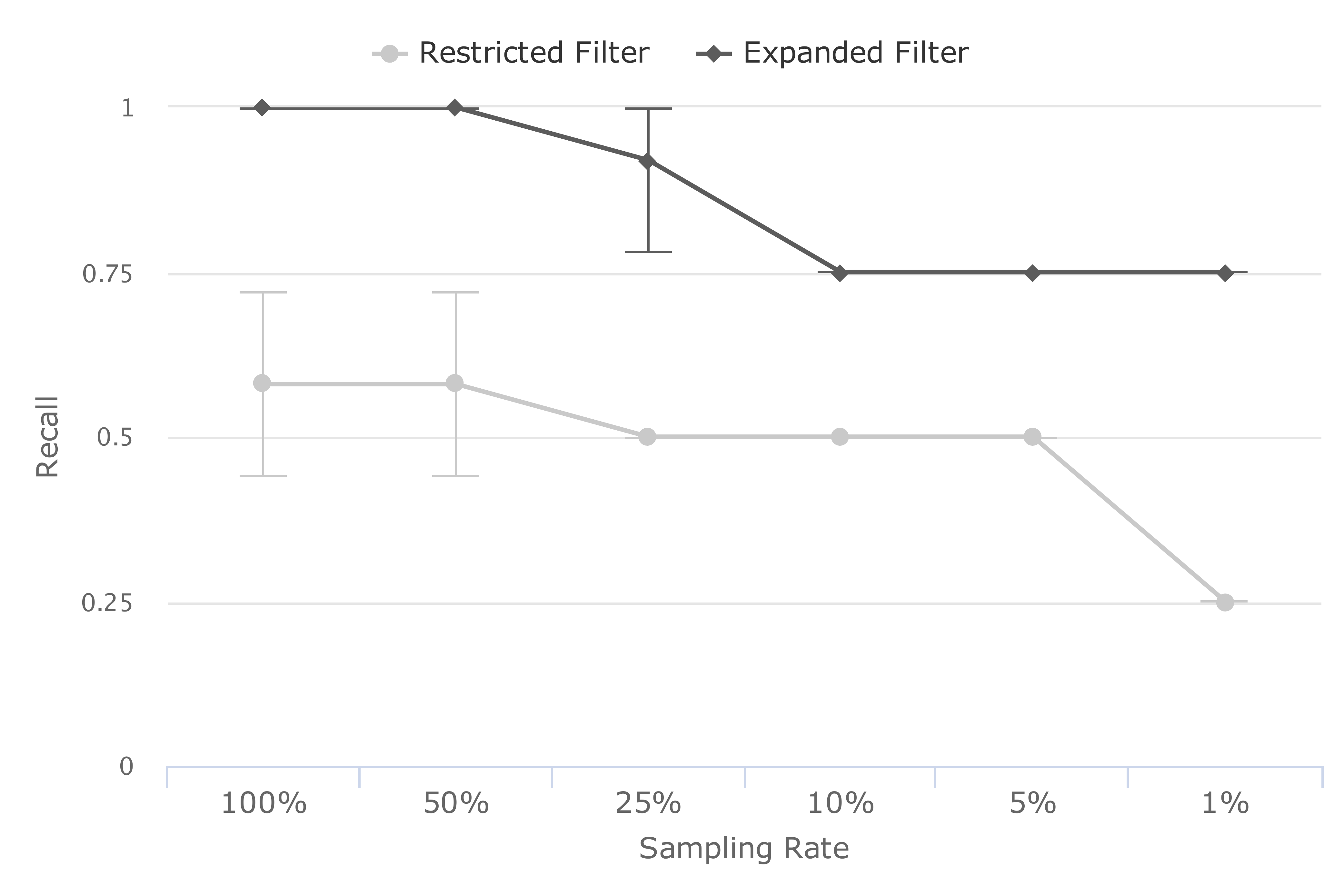}
        \caption{Pet Clinic}
        \label{fig:recall-petclinic}
    \end{subfigure}
    \begin{subfigure}{\linewidth}
        \includegraphics[width=\linewidth]{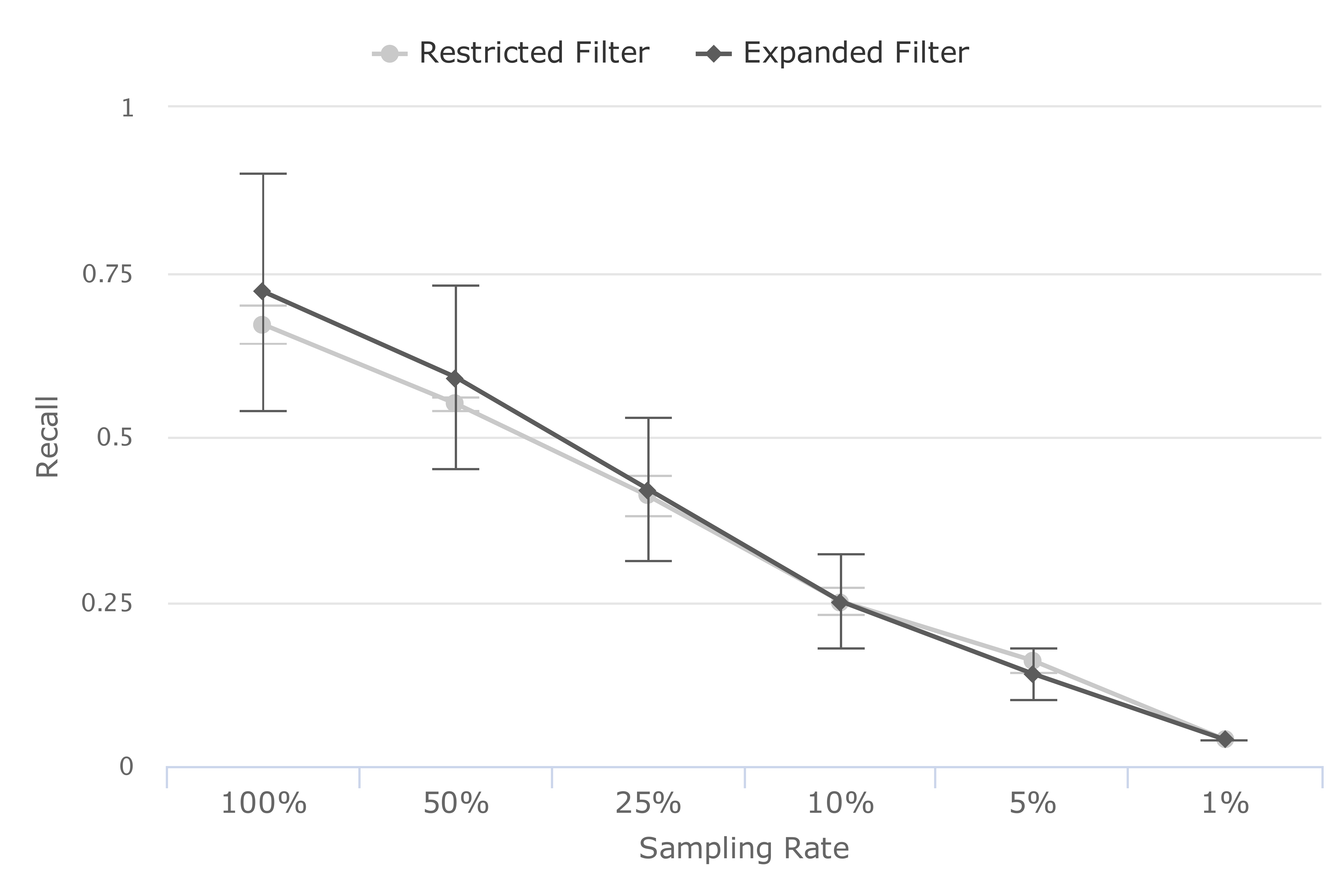}
        \caption{Shopizer}
        \label{fig:recall-shopizer}
    \end{subfigure}
    \caption{RQ2: Recall by Sampling Rate. Recall obtained for each combination of filter (Restricted Filter and Expanded Filter) and sampling rate.}\label{fig:resultsRQ2}
\end{figure}

Although the number of selected methods to be cached using Tigris is smaller in comparison to APLCache monitoring, we observe that this number changes according to the filtering and sampling configuration. For all applications, restricting the filter and decreasing the sampling rate may significantly reduce the number of caching opportunities. The amount of information collected is reduced, and thus the accuracy of APLCache may be compromised.

None of the combinations of filters and sampling rates can identify all the opportunities initially identified by the original APLCache. However, the throughput is still improved due to the filtered monitoring guided by the coarse-grained phase of our approach. The absolute number of hits varies according to the number of cacheable opportunities. The lower number of hits follows reduced cacheable opportunities. The hit ratio remains almost the same along with all the simulations, and small variations are due to the variations in cacheable opportunities found at different monitoring cycles.

We also analyze the effectiveness achieved with each set of execution traces obtained using Tigris with the different filters and sampling rates. Although improving the overall performance is important, our key goal is to record the traces that are actually those needed for the analysis made by APLCache. Consequently, we consider as ground-truth the set of caching opportunities identified with full monitoring and compare it to the sets obtained with the various Tigris configurations. We measure this using typical classification performance metrics, namely precision, recall and f-measure. Results are presented in Table~\ref{tab:results} (columns Pr., Recall and F1, respectively).

As can be seen, the precision---for both filters and sampling rates---is 1.00, that is, there are no false positives. This means that the subset of methods to be monitored in the second phase allows APLCache to identify caching opportunities correctly. These results, however, do not hold for recall. The recall decreases when the filter is more restricted and the sampling rate is lower. Filters can cause relevant methods (those that should be identified as a cacheable opportunity) to be not monitored in the second phase. Consequently, they are not traced and considered for caching. The low sampling rate, in turn, can lead to samples that are not representative of the population of the method calls. The variation of recall across the different Tigris configurations can be seen in Figure~\ref{fig:resultsRQ2}. It shows that the filter has a larger effect on recall than the sampling rate.

The restricted filter, for CloudStore and PetClinic, is the main cause for a low recall. It can be seen that even with sampling rates higher than 5\%, the recall remains quite similar or even the same. For Shopizer, although the recall with 100\% is not high (0.66), it decreases as the sampling rate decreases. Both the filter and the sampling rate, therefore, affect the recall. Although the expanded filter leads to false negatives, it achieves high recall, up to 1.00 (i.e.\ all the cacheable opportunities were found). Increasing the sampling rate improves the recall for all applications, as more information is provided for APLCache to analyze. For CloudStore and PetClinic, the highest recall can be achieved even with sampling rates lower than 100\%. Shopizer potentially has results different from the other applications because it is larger and thus has a broader range of methods. In addition, the margin of error of the recall, observed in the different monitoring cycles, demonstrated to be high for the expanded filter. It demonstrates how the workload variations can impact the number of methods selected for fine-grained monitoring and the number of cacheable opportunities found at each monitoring cycle by APLCache.

\begin{framefloat}
\noindent \textbf{RQ2: Findings.} The relevance filter and sampling rate of Tigris can reduce the number of identified caching opportunities and, consequently, the number of hits. The hit ratio, however, remains almost the same (0.91--0.98). Therefore, lower performance gains are obtained due to less cached opportunities.
Tigris leads to no false positives considering APLCache and the target applications, thus achieving a precision of 1.00. However, recall can decrease due to the used relevance filter and low sampling rates, being larger the effect of the filter.
\end{framefloat}

\subsubsection{RQ3. How does Tigris cope with workload variations over time?}

To evaluate how changes in the workload impact the methods selected by the first phase of the proposed approach, we inspected the subset of methods selected by the first phase of Tigris at the end of each monitoring cycle. Obtained results are presented in Table~\ref{tab:results-adaptation}.
We first observe that the number of selected methods for fine-grained monitoring changed on every cycle for all the applications. These changes were expected since the relevance criteria are domain-neutral, and the metrics used to analyze them do not rely on pre-defined thresholds or assumptions regarding the workload. As a consequence, the selected methods are based on the application's emerging behavior. In addition, the results show that bigger applications (i.e.\ with more methods and navigation paths) such as CloudStore and Shopizer, tend to have significant differences among monitoring cycles. As Petclinic has fewer navigation paths to be executed, the workload variations do not affect much the relevance-based selection. Still, all Petclinic monitoring cycles resulted in changes in the selection. This demonstrates the ability of our approach to adapt to workload variations.
For all the applications, the restricted filter results in less relevant methods being selected for detailed monitoring than the expanded filter.


\begin{table}
\renewcommand{\tabcolsep}{1.3mm}
\centering
\caption{Simulation Results. Results in terms of changes in the relevance evaluation and selected methods to monitor for each application with restricted and expanded filters. After each monitoring cycle a snapshot of the coarse-grained selection is taken, reporting the amount of selected methods in that cycle (Selected), the overall difference from the last cycle (Difference), including the specific amount of additions and exclusions.}
\label{tab:results-adaptation}
\begin{tabular}{llrrr}
\toprule
& \textbf{Monitoring} & \textbf{Cycle} & \textbf{Selected} & \textbf{Difference} \\ \midrule 

\multicolumn{1}{l}{\multirow{6}{*}{\rotatebox[origin=c]{90}{\textbf{Cloud Store}}}} 
 & \multirow{3}{*}{Restricted Filter} & 1 & 8 & \\
 &  & 2 & 14 & +6 (+8/-2) \\
 &  & 3 & 13 & -1 (0/-1) \\ \cmidrule(l){2-5}
 
 & \multirow{3}{*}{Expanded Filter} & 1 & 36 & \\
 &  & 2 & 24 & -12 (+0/-12) \\
 &  & 3 & 20 & -4 (+1/-5) \\ \midrule

\multicolumn{1}{l}{\multirow{6}{*}{\rotatebox[origin=c]{90}{\textbf{Petclinic}}}}  
 & \multirow{3}{*}{Restricted Filter} & 1 & 4 & \\
 &  & 2 & 5 & +1 (+2/-1) \\ 
 &  & 3 & 4 & -1 (+1/-2) \\ \cmidrule(l){2-5}
 
 & \multirow{3}{*}{Expanded Filter} & 1 & 14 & \\
 &  & 2 & 13 & -1 (+0/-1) \\ 
 &  & 3 & 12 & -1 (+0/-1) \\ \midrule

\multicolumn{1}{l}{\multirow{6}{*}{\rotatebox[origin=c]{90}{\textbf{Shopizer}}}}
 & \multirow{3}{*}{Restricted Filter} & 1 & 33 & \\ 
 &  & 2 & 45 & +12 (+13/-1) \\ 
 &  & 3 & 38 & -7 (+2/-9) \\ \cmidrule(l){2-5}
 
 & \multirow{3}{*}{Expanded Filter} & 1 & 76 & \\ 
 &  & 2 & 55 & -21 (+6/-27) \\ 
 &  & 3 & 62 & +7 (+11/-4) \\ \bottomrule
\end{tabular}
\end{table}

\begin{framefloat}
\noindent \textbf{RQ3: Findings.} Because the proposed relevance criteria and metrics are domain-neutral and do not rely on pre-defined thresholds or assumptions, Tigris can adapt to different workloads, identifying relevant methods according to the application's emerging behavior.
\end{framefloat}

\subsection{Threats to Validity}

We now analyze the threats to the validity of our empirical evaluation. First, the performance impact of monitoring highly depends on workloads. Although we do not make any assumptions regarding the workload and rely on the randomness added to the tests, the workload used in the experiments may not be representative enough to be generalized. Nevertheless, our approach does not depend on a particular workload and can find relevant traces with any pre-specified workload, which may evolve in real-world scenarios. Therefore, even if the workload changes substantially and initial relevant methods are no longer useful, our approach can adapt itself, automatically discarding outdated monitoring configurations and discovering a new set of relevant execution traces.

Second, our evaluation is limited to one goal of monitoring (i.e.\ application efficiency in terms of performance) and only one monitoring-based approach (i.e.\ a caching technique). Therefore, the results may not be generalizable. To address this threat, we selected three open-source target applications, with different sizes and domains, implemented by different developers. In addition, we provide a wide variety of the tunable parameters for adaptive monitoring (i.e.\ the relevance filter and sampling rate) and compare the results against a baseline. We acknowledge that all of the threats mentioned above may require several evaluations concerning multiple systems of different sizes, users, traces, workloads, and other environmental conditions that should be addressed as part of future work.

\subsection{Limitations}

Providing a solution for effective execution tracing requires dealing with many challenges other than those addressed in this paper, such as overhead management, sampling gaps, and defining appropriate criteria, metrics and sampling rate to achieve the goal of monitoring. Consequently, although our approach makes substantial advances towards software monitoring, there are challenges that remain open. These correspond to limitations of our work, which are discussed as follows.

Both filtering and sampling techniques, which are the basis of our approach, are prone to introducing gaps in the output trace. Although filtering is based on the goal of monitoring, given as input of the approach, the sampling strategy follows a sampling rate, leading to uncertainty in the monitoring result in terms of representativeness. To quantify such uncertainty, a possible approach is learning statistical models of the monitored traces and use them to fill in sampling-induced gaps and then to compute the probability that the property of interest is satisfied or violated~\cite{Bartocci2018}.

In addition, choosing a sampling rate is not an easy task and depends on the representativeness needed and also the supported overhead. A low sampling rate may give a lousy precision, and a high sampling rate may generate a considerable amount of useless data and overhead. In addition, for some large-scale distributed systems, 1\% of tracing data might be quite demanding for analyzing and decision making and might still provide a large overhead for the system. To deal with this situation, we can adopt an adaptive sampling, which dynamically adjusts the sampling rate by observing the impact of sampling rates on the overall computational resource usage.

The coarse-grained monitoring demands to instrument all method calls, and thus a minimal but additional overhead is incurred due to this per-instruction instrumentation before the filtering and sampling decision making. To reduce even more the overhead of the approach, we can also apply a dynamic sampling strategy into the coarse-grained monitoring.

Although the framework supports the use of already implemented estimations of metrics, which were identified and conceived based on the investigation of existing monitoring-based approaches, our approach does not cover the challenge of deciding which estimate or criteria are appropriate to specific goals of monitoring. In fact, tunable parameters for adaptive monitoring (in terms of relevance criteria and sampling rate) create a configuration space and as a result expose a secondary problem of finding appropriate values for such configuration options. Thus it is not in the scope of this paper. However, the results of our foundational study includes a list of relevance criteria, an occurrence-based association between groups of goals and relevance criteria, and examples of goals and metrics used by the surveyed approaches. This information provides support to specify or at least reduce the configuration space created by our approach. In addition, although the available criteria, goals, and metric estimations achieved good results in our experiment, they may not fit well in all the domains and workloads. To solve this problem, Tigris was designed to be extensible and flexible, providing interfaces that can be used to customize and change how metrics are calculated.

Regarding the classification of data into groups, we do not deal with possible outliers that may appear due to the transient behavior of the application, such as an increase in the execution time of a method given the high level of concurrency. It can be addressed in the future with enhanced statistical analysis and filtering of outliers.
In addition, some tests for normality may be not sensitive enough given the sample size and the property of the data. Ideally, testing for normality should be executed and interpreted alongside histograms, QQ-plots, and skewness and kurtosis values. To solve this problem, the framework can be evolved in such a way that small parts of the monitoring phases could be customized by users, such as how to classify data.

The coarse-grained monitoring data is stored in memory, and despite its low memory usage, it may reach an imposed limit if kept forever. In our evaluation, this was not an issue. However, this can be configured in the form of a time frame and added as an additional parameter of the framework. We also need to understand how we could combine sliding windows of monitoring to avoid losing historical information.

Although the framework is adaptive as it can vary the selected list of relevant methods according to the application’s emerging behavior, it is necessary to specify when this adaptation should be triggered. In our evaluation, it uses a two-minute interval. However, choosing the most appropriate interval to update the list of relevant methods involves a trade-off between collecting enough lightweight information about the application behavior to reach good decisions and changing the list fast enough to keep it in sync with the application behavior and collect more relevant traces. To solve this problem, future work can provide an adaptive triggering strategy, which detects significant variations in the workload characteristics or degradation in the quality of traces being collected, according to the goal of monitoring.

\section{Conclusion}
\label{sec:conclusion}

Monitoring is a crucial step to support software engineering tasks, such as software adaptation. However, developing effective monitoring tools demands a significant effort, considering the existing challenges such as the monitoring overhead, which can impact the performance of the system. In this paper, we presented a systematic literature review that allowed us to identify relevance criteria and metrics that can be used for monitoring applications. We also analyzed existing monitoring-based approaches in terms of generality, scalability, and adaptability. Most of the existing solutions are application-specific, with principles to monitor data that are not generalizable. For scaling monitoring, typical approaches sample or filter collected traces or limit the monitoring to high-level events. Adaptations in existing solutions are focused on controlling the monitoring overhead or increasing the trace reliability for a specific purpose.

Based on the results of the literature review, we derived a domain-specific language (DSL), named TigrisDSL, which allows the creation of monitoring filters by means of high-level relevance criteria. These relevance filters can be used to guide monitoring components to collect a set of relevant traces that are analyzed to achieve the goal of monitoring. This DSL was incorporated into a proposed two-phase approach for low-impact monitoring of execution traces. The approach is domain-neutral and can be instantiated to collect relevant traces for different domains and purposes.

The first phase of our approach is based on obtaining lightweight metrics from the system's execution. Then, from time to time, it evaluates the collected metrics according to a set of relevance criteria specified in TigrisDSL. The second phase filters and samples executions that are relevant to the goal of monitoring, thus achieving a reduced monitoring overhead. We implemented the proposed approach as a Java-based framework, called Tigris, and instantiated it as the monitoring component of an existing monitoring-based approach, which identifies caching opportunities at runtime by analyzing method calls. The results show that, while full monitoring applications causes an overhead of 22.6\%--29.9\%, Tigris reduces this overhead to 2.9\%--24.7\%. Considering the overall application performance, Tigris can improve it up to 27.4\%. In comparison to the cacheable opportunities identified with full monitoring, Tigris manages to achieve a precision of 1.0. However, recall largely varies in terms of the filter and sampling rate adopted. Lastly, given the proposed relevance criteria and that metrics are domain-neutral, Tigris can adapt to different workloads, identifying relevant methods according to the application's behavior.

Tigris was conceived and implemented in a flexible and extensible way, paving the way to cover more unaddressed monitoring challenges in the future. Although the framework provides a Java implementation of our approach, our two-phase monitoring proposal is generic and can be implemented on any programming language. Future work involves addressing the framework limitations and enhancing it with adaptive capabilities, such as taking the effectiveness of monitoring as feedback to adjust the framework parameters, such as increasing the sampling rate or expanding the set of filtered methods. In addition, we intend to investigate techniques to adapt the sampling rate according to the system's load that ensures a representative sample with an acceptable performance impact.

\section*{Acknowledgments}
Ingrid Nunes thanks for CNPq grants ref. 313357/2018-8 and ref. 428157/2018-1. This study was financed in part by the Coordena\c{c}\~{a}o de Aperfei\c{c}oamento de Pessoal de N\'{i}vel Superior - Brasil (CAPES) - Finance Code 001.

\bibliographystyle{model1-num-names}
\bibliography{references}
\end{document}